\documentclass{llncs}

\usepackage{wrapfig,lipsum,booktabs}
\usepackage{url}
\usepackage{eurosym}
\usepackage{multicol}
\usepackage{epsfig}
\usepackage{subfigure}
\usepackage{graphicx}
\usepackage{commath}
\usepackage{color, colortbl}
\usepackage{algorithm}
\usepackage[noend]{algorithmic}
\usepackage{amsmath}
\usepackage{paralist}
\usepackage{tikz}
\usepackage{epstopdf}
\usetikzlibrary{decorations.pathreplacing}
\floatname{algorithm}{Algorithm}

\newcommand{\argmax}{\mathop{\mathrm{arg\,max}}}

\newtheorem{lem}{\bf Lemma}
\newtheorem{Pro}{\bf Proposition}
\newtheorem{thm}{\bf Theorem}
\newtheorem{Cor}{\bf Corollary}

\usepackage{colortbl}
\definecolor{Gray}{gray}{0.90}

\floatname{algorithm}{Algorithm}

\newif\ifExtendedVersion
\ExtendedVersiontrue

\begin{document}

\allowdisplaybreaks

\setlength{\textfloatsep}{3pt plus 1.0pt minus 2.0pt}
\setlength{\intextsep}{3pt plus 1.0pt minus 2.0pt}

%
\title{An Economic Study of the Effect of Android Platform Fragmentation on Security Updates}

\author{Sadegh Farhang\inst{1}, Aron Laszka\inst{2}, and Jens Grossklags\inst{3}}
\institute{Pennsylvania State University \and University of Houston \and Technical University of Munich\\ \email{farhang@ist.psu.edu, alaszka@uh.edu, jens.grossklags@in.tum.de}}


\titlerunning{Economic Study of the Effect of Android Platform Fragmentation}
\authorrunning{Farhang, Laszka and Grossklags}

\maketitle

\begin{abstract}
Vendors in the Android ecosystem typically customize their devices by modifying Android Open Source Project (AOSP) code, adding in-house developed proprietary software, and pre-installing third-party applications. However, research has documented how various security problems are associated with this customization process.

We develop a model of the Android ecosystem utilizing the concepts of game theory and product differentiation to capture the competition involving two vendors customizing the AOSP platform. We show how the vendors are incentivized to differentiate their products from AOSP and from each other, and how prices are shaped through this differentiation process. We also consider two types of consumers: security-conscious consumers who understand and care about security, and na\"ive consumers who lack the ability to correctly evaluate security properties of vendor-supplied Android products or simply ignore security. It is evident that vendors shirk on security investments in the latter case. 

Regulators such as the U.S. Federal Trade Commission have sanctioned Android vendors for underinvestment in security, but the exact effects of these sanctions are difficult to disentangle with empirical data. Here, we model the impact of a regulator-imposed fine that incentivizes vendors to match a minimum security standard. Interestingly, we show how product prices will decrease for the same cost of customization in the presence of a fine, or a higher level of regulator-imposed minimum security.
\end{abstract}



\section{Introduction}
\label{sec:Intro}
%
%
Android, the mobile operating system released under open-source licenses as the Android Open Source Project (AOSP), has the largest market share among smartphone platforms worldwide with more than one billion active devices~\cite{MarketShare}. 
Due to the openness of the platform, vendors
and carriers can freely customize features to differentiate their products from their competitors. This differentiation includes customizing the hardware, but there is also a substantial fragmentation in the software packages utilized in the Android ecosystem~\cite{han2012understanding,Fragment}.

The fragmentation of the software base available from various vendors is due to various customization steps, including the modification of the open source Android codebase as well as the addition of proprietary software. Product differentiation may benefit consumers by providing Android devices for sale that better match consumer tastes, and may also benefit businesses by helping them to sidestep intense price competition of homogeneous product markets \cite{tirole1988theory}.

However, we also observe that Android platform fragmentation is associated with a number of security challenges~\cite{Thomas15,wu2013impact,zhou2014peril}. For example, Wu et al. showed that a large proportion of security vulnerabilities in the Android ecosystem are due to vendor customization. They calculated that this proportion is between $64\%$ to $85\%$ for different vendors~\cite{wu2013impact}. Similarly, Zhou et al. showed how customized drivers for security-sensitive operations on Android devices available by different vendors often compare unfavorably to their respective counterparts on the official Android platform~\cite{zhou2014peril}. Thomas et al. provided evidence for the substantial variability of security patch practices across different vendors and carriers \cite{Thomas15}. Using a dataset about over 20,000 Android devices, they showed that on average over 87\% of the devices are exposed to at least one of 11 known critical (and previously patched) vulnerabilities.\footnote{Further compounding the problem scenario is how third-party apps targeting outdated Android versions and thereby disabling important security changes to the Android platform cause additional fragmentation~\cite{mutchlertarget}.} 


The Android ecosystem fragmentation and the associated security problems have caused consumer protection agencies to intervene in the marketplace. In 2013, the Federal Trade Commission (FTC) charged a leading vendor because it ``failed to employ reasonable and appropriate security practices in the design and customization of the software on its mobile devices'' \cite{FTC13}. The case was settled and the vendor was required to ``establish a comprehensive security program designed to address security risks during the development of new devices and to undergo independent security assessments every other year for the next 20 years'' \cite{FTC13}. Not observing significant improvements in the Android ecosystem, the FTC recently solicited major vendors to provide detailed information about their security practices including what vulnerabilities have affected their devices as well as whether and when the company patched those vulnerabilities~\cite{FTC}. 

In this paper, we propose a product differentiation model that captures key facets of the Android ecosystem with a focus on the quality of security. We consider multiple competing vendors, who can customize Android for their products in order to differentiate themselves from their competitors. We consider both security-conscious consumers, who value security quality, and na\"ive consumers, who do not take security issues into consideration when they make adoption choices. When consumers are na\"ive, vendors do not have any incentives to address security issues arising from the customization. In order to incentivize investing in security, a regulator may impose a fine on vendors that do not uphold a desired level of security. We show that fines can achieve the desired effect, and we study how they impact the competitive landscape in the Android ecosystem. 

\textbf{Roadmap}. In Section~\ref{sec:BackG}, we provide background on Android customization and the associated security challenges. Section~\ref{sec:SysMod} presents the economic model on Android customization. We analyze the model without a fine in Section~\ref{sec:Ana} and consider how to calculate the parameters in our model in Section~\ref{sec:Num}. We extend the model to the case with a regulator-imposed fine in Section~\ref{sec:SysModFine}. We support our analysis with numerical results in Section~\ref{sec:NumiLL}. 
We conclude in Section~\ref{sec:Conclu}.

\section{Background}
\label{sec:BackG}


\textbf{Customization:} One approach to measure the level of customization by vendors is \textit{provenance} analysis~\cite{wu2013impact}, which studies the distribution and origin of apps on Android devices. There are mainly three sources of app origins on Android devices: (1) AOSP: apps available in the default AOSP that, however, can be customized by a vendor; (2) Vendor: apps that were developed by that vendor; and (3) Third-party: apps that are not in AOSP and were not developed by the vendor. 

Table~\ref{tab:Prov} summarizes the published findings of a provenance analysis of five popular vendors: Google, Samsung, HTC, Sony, and LG \cite{wu2013impact}. 
The authors found that on average $18.22\%$, $64.41\%$, and $17.38\%$ of apps originate from AOSP, vendors, and third parties, respectively. Further, the number of apps and lines of code (LoC) associated with the devices are increasing with newly released versions. Likewise, the complexity of the baseline AOSP is increasing over time~\cite{wu2013impact}. 

\begin{table*}[h!]
	\centering
	\tiny
	\begin{tabular}{|c|c|c|c|c|c|c|c|c|c|c|} \hline
		\multicolumn{5}{|}{} & \multicolumn{2}{|c}{\textbf{AOSP}} & \multicolumn{2}{|c}{\textbf{vendor}} & \multicolumn{2}{|c|}{\textbf{3rd-party}} \\ \hline
		\textbf{Vendor} & \textbf{Device} & \textbf{Version and Build$\#$} & \textbf{$\#$apps} &  \textbf{$\#$LOC} & \textbf{$\#$apps} & \textbf{$\#$LOC} & \textbf{$\#$apps} & \textbf{$\#$LOC} &  \textbf{$\#$apps} &  \textbf{$\#$LOC} \\ \hline
		Samsung & Galaxy S2 & 2.3.4; 19100XWKI4 & 172 & 10M & 26 & 2.4M & 114 & 3.5M & 32 & 4.1M \\ \hline
		Samsung & Galaxy S3 & 4.0.4; 19300UBALF5 & 185 & 17M & 30 & 6.3M & 119 & 5.6M & 36 & 5.3M \\ \hline		
		HTC & Wildfire S & 2.3.5; CL362953 & 147 & 9.6M & 24 & 2.7M & 94 & 3.5M & 29 & 3.3M \\ \hline
		HTC & One X & 4.0.4; CL100532 & 280 & 19M & 29 & 4.7M & 190 & 7.3M & 61 & 7.5M \\ \hline
		LG & Optimus P350 & 2.2; FRG83 & 100 & 6.1M & 27 & 1.1M & 40 & 0.6M & 33 & 4.4M \\ \hline
		LG & Optimus P880 & 4.0.3; IML74K & 115 & 12M & 28 & 3.1M & 63 & 3.2M & 24 & 5.6M \\ \hline
		Sony & Xperia Arc S & 2.3.4; 4.0.2.A.0.62 & 176 & 7.6M & 28 & 1.1M & 123 & 2.6M & 25 & 3.8M \\ \hline
		Sony & Xperia SL & 4.0.4; 6.1.A.2.45 & 209 & 10M & 28 & 1.8M & 156 & 4.1M & 25 & 4.7M \\ \hline
		Google & Nexus S & 2.3.6; GRK39F & 73 & 5.2M & 31 & 1M & 41 & 2.8M & 1 & 1.3M \\ \hline
		Google & Nexus 4 & 4.2; JOP40C & 91 & 15M & 31 & 2.5M & 57 & 12M & 3 & 1.1M \\ \hline
	\end{tabular}
	\caption{Provenance analysis~\cite{wu2013impact}.}
	\label{tab:Prov}
\end{table*}

One of the challenges in this fragmented ecosystem is 
the security risk that arises from the vendors' and carriers' customization to enrich their systems' functionality without fully understanding the security implications of their customizations. In this paper, our focus is on security issues resulting from such customization. Hence, we focus on this issue in the following subsection.


\textbf{Security Impact of Customization:} The problems related to security aspects of Android customization are mainly due to vendors' change of critical configurations. These changes include altering security configurations of Linux device drivers and system apps, etc. One approach for better understanding the effect of customization is to compare security features of different Android devices with each other, which is called differential analysis.
Aafer et al. proposed a number of security features to take into account~\cite{aafer2016harvesting}.
First, \textit{permissions} which protect data, functionalities, and inner components can be analyzed. In Android, we have four level of permissions: \textit{Normal}, \textit{Dangerous}, \textit{Signature}, \textit{SystemOrSignature}. The goal of differential analysis is to find a permission with a different (and typically lower) level of protection on some devices. Second, group IDs (\textit{GIDs}) are another feature to take into account. Some lower-level GIDs are given Android permissions, which could potentially be mapped into a privileged permission due to customization. Protected broadcasts sent by system level processes are a third important security feature. Due to customization, some protected broadcasts could be removed and, as a result, apps can be triggered by not only system-level processes but also by untrusted third-party apps. 

By comparing these security features, Aafer et al. found that the smaller the vendor is, the more significant inconsistencies are observable for the different security features. One interpretation is that the cost of investment in security is too high for those vendors (e.g., hiring of security experts). The results also imply that different vendors invest in security to different degrees. 

Research aiming to understand Android customization is clearly demonstrating that customization is a pervasive feature in Android, and this is associated with a wide variety of security challenges and vulnerabilities. Further, we are unaware of any research that provides evidence for security improvements resulting from customization, which outweighs the aforementioned risks. At the same time, research is missing that aims to understand the economic forces associated with the customization process which is the objective of our work.

\section{Model Definition}
\label{sec:SysMod}
In this section, we propose our baseline model
in the tradition of \textit{game theory} and the \textit{theory of product differentiation} \cite{tirole1988theory,hotelling1990stability}. Our model considers three types of entities: (1) AOSP, (2) vendors, such as Samsung or LG, and (3) consumers. 

\textbf{AOSP:} Google, the developer of Android, provides monthly security updates for its devices and for base Android. However, other vendors have to adjust AOSP security updates for their Android devices because of their customization. Further, customization may also introduce new security vulnerabilities. 

To incorporate these effects into our model, we assume that a customized version of Android can be represented by a point on the segment $[0, 1]$. Our analysis could be extended to multidimensional customizations in a straightforward way, but we assume one dimension for ease of presentation, since our focus is on the relative level of customization rather than its direction. Moreover, the location of each Android customization is independent of objective measures of product quality. 
In other words, we map the features of a mobile device to a point on the segment $[0, 1]$ to quantify its difference (e.g., percentage of customized code) from the base version of Android provided by AOSP. In our model, $Z_A$ denotes the point corresponding to the base version of AOSP.  Since AOSP aims to provide a base version that maximizes the market share of Android, it provides a version that can attract the widest range of consumers. 
Hence, in the numerical analysis, we assume that AOSP is in the middle, i.e., $Z_A=0.5$.


\textbf{Vendor:} There are multiple vendors selling Android devices. Carriers can sell the vendors' Android devices with their own prices and customizations too. Here, we will use the term ``vendor'' to refer to both vendors and carriers. The price and the market share of the device sold by vendor $i$ are denoted by $p_i$ and $D_i$, respectively.  Further, $q_i$ denotes the security quality of patches delivered by vendor $i$. We assume that $p_i \geq 0$ (product prices are non-negative) and $q_i \geq 0$ (security quality is represented by a non-negative number) for every vendor $i$.
Similar to the AOSP base version, 
a point $z_i \in [0,1]$ represents the customization of the Android version of vendor $i$. 

We consider two types of costs for customization. 
First, through customization, the vendor makes its product different from what Google has developed in AOSP. Hence, the vendor incurs development cost, which is related to the degree of customization. Here, we model this cost as a convex quadratic function of the difference between the vendor's position and the positions of the AOSP base version. 
Second, the security related cost of a vendor depends not only on the quality and frequency of security updates provided by the vendor, but also on the difference due to customization. Vendors receive security patch updates from AOSP, but due to customizations, vendors need to adapt these security patches before distribution. Often, vendors degrade the quality or frequency of security patches in order to save development and distribution costs~\cite{Thomas15}. Hence, the security-related cost is affected by both the customization level and the security quality. In our model, we employ a convex quadratic function to capture how the security cost of vendor $i$ depends on $q_i$.  
The utility of vendor $i$ is equal to:
\begin{equation}
\pi_i = p_i D_i - C_i  \left(z_i - Z_A \right)^2 - S_i q_i^2 \left(z_i - Z_A \right)^2,
\label{eq:venforUtilPre}
\end{equation}
where $C_i$ and $S_i$ are constants representing cost per unit of customization and security quality, respectively. 
Here, we focus on security issues resulting from Android customization rather than security-related cost of AOSP explicitly.

We have considered quadratic functions for the cost terms,
which is a common assumption for modeling customization costs, e.g., see~\cite{cavusoglu2007selecting} and \cite{dewan2003product}. The quadratic cost function captures the fact that the cost of customization increases as the customization increases. In a similar way, with an increase in the cost of customization or the quality of security, the security cost resulting from customization increases. It would be possible to use any functional form with increasing marginal cost, such as an exponential cost function, which would lead to the same qualitative results as the ones presented here. 

We also consider the quality of security patch updates provided by AOSP, denoted by $Q$, to be an exogenous parameter in our model, which applies to all vendors in the same way.
Note that we observe that in practice, vendors virtually never provide better security quality. Further, we are primarily interested in studying the effect of customization on security; hence, we will not consider vendors implementing additional security measures that are independent of customization. Hence, we assume that the value of $q_i$ is upper bounded by $Q$.
 


\textbf{Consumers:} Consumers choose mobile devices primarily based on prices and how well the devices match their preferences, but they may also consider security quality.
A consumer's preference, similar to a vendor's customization, can be represented by a point $x$ in $[0,1]$. Consumers' preferences for smartphone selection are heterogeneous and we assume that the consumers' preferences are distributed uniformly in $[0,1]$. 
We consider security-conscious consumers who take security into account when choosing their product. 
The utility of consumer~$j$ for choosing Android type $i$ given that consumer $j$ is at $x_j$ is:
\begin{equation}
u_j^i = \beta q_i - p_i - T\left(x_j-z_i\right)^2,
\label{eq:ConPref2}
\end{equation}
where $T$ represents the consumer's utility loss for one unit of difference between its preference and the location of the product, which we call \textit{customi\-zation-importance}. Similarly, $\beta$ represents the consumer's utility gain for one unit of security quality, which we call \textit{security-importance}. Na\"ive consumers, who do not understand or care about security quality, can be modeled by letting $\beta = 0$. 

Our utility for consumers is in agreement with literature in economics~\cite{d1979hotelling}. It is common to consider quadratic term in economics to model utility. 

\begin{figure}[h]
	\centering
	\begin{tikzpicture}[thick, y=0.4cm, x=0.6cm]
	\draw [very thick] (0,0) -- (10,0);
	\draw [very thick] (0,0.3) -- (0,-0.3);
	\draw [very thick] (10,0.3) -- (10,-0.3);
	\node at (0,-1) {$0$};
	\node at (10,-1) {$1$};
	\draw [fill] (2.5,0) circle (0.07cm);
	\node at (2.5,-1) {Vendor $1$};
	\draw [fill] (5,0) circle (0.07cm);
	\node at (5,-1) {AOSP};
	\draw [fill] (7.5,0) circle (0.07cm);
	\node at (7.5,-1) {Vendor $2$};
	\draw [decorate, decoration={brace,amplitude=5pt}, yshift=0.2cm]
	(0,0) -- (2.5,0) node [midway, above, yshift=0.2cm] {$a$}; 
	\draw [decorate, decoration={brace,amplitude=5pt}, yshift=0.2cm]
	(7.5,0) -- (10,0) node [midway, above, yshift=0.2cm] {$b$};  
	\end{tikzpicture}
	\caption{Location of vendor $1$ and vendor $2$.}  
	\label{fig:SystemModel}
\end{figure}
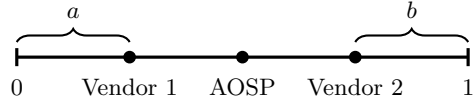

\textbf{Game Formulation}: For tractability, we consider a two-player game between vendor $1$ and vendor $2$ without any other vendors.\footnote{While we restrict our model to two vendors, we are aware that in practice, there are more than two vendors competing with each other. However, we believe that similar to classic economic studies with two companies in the context of product differentiation, our model provides a meaningful understanding of the customization in the Android ecosystem and of security quality.} 
In our analysis, we assume that vendors are on different sides of AOSP. Further, we let $ a = z_1$ and $ 1-b = z_2$.  Without loss of generality, we assume that $0 \leq a \leq 1-b \leq 1$. Figure~\ref{fig:SystemModel} shows the location of vendor 1 and vendor $2$. 

The utilities of vendors $1$ and $2$ are then as follows:
\begin{align}
\pi_1 = p_1 D_1 - C_1 \left(a - Z_A\right)^2 - S_1 q_1^2\left( a - Z_A\right)^2,
\label{eq:venforUtil1} \\
\pi_2 = p_2 D_2 - C_2 \left(1 - b - Z_A\right)^2 - S_2 q_2^2\left( 1 - b - Z_A\right)^2.
\label{eq:venforUtil2}
\end{align}

To calculate the Nash equilibrium, we need to define the stages of the game, i.e., the order in which the two players choose their prices, locations, and security levels. For our analysis, we consider the following stages:

\begin{compactitem}
	\item[$\bullet$] \textbf{Stage 1}: Both vendors simultaneously choose their location parameters $a$ and $b$. They also choose their level of security quality, i.e.,  $q_1$ and $q_2$.
	
	\item[$\bullet$] \textbf{Stage 2}: Both vendors simultaneously choose their prices $p_1$ and $p_2$.

\end{compactitem}

The reason is that a vendor freely modifies the AOSP code base, adds its developed proprietary software, and installs a diverse set of third-party apps to customize its device. These changes, however, result in the change of critical configurations leading to security issues~\cite{Thomas15,wu2013impact,zhou2014peril}. Therefore, it is reasonable to consider that the customization and security quality effort happen at the same stage. Then, by taking into account its effort in customization and security quality, the vendor chooses its price. 
We use backward induction to solve our game. 
First, we consider stage 2 and calculate the price Nash equilibrium for given locations and quality. Then, we consider stage 1 and calculate the location and quality equilibrium assuming a price equilibrium in stage 2. 

Table~\ref{tab:symbols} shows a list of the symbols used in our model.

\begin{table}[h!]
	\centering
	\caption{List of Symbols}
	\label{tab:symbols}
	\begin{tabular}{|c|l|}
		\hline
		Symbol & Description \\ 
		\hline
		$Z_A$ & Point corresponding to AOSP \\
		\rowcolor{Gray} $D_i$ &  Market share of vendor $i$\\
		$z_i$ &  Customization of the Android version of vendor $i$\\
		\rowcolor{Gray} $p_i$ &  Price of vendor $i$\\
		$q_i$ &  Security quality of patches delivered by vendor $i$\\
		\rowcolor{Gray} $S_i$ &  Cost per unit of security quality\\
		$C_i$ &  Cost per unit of customization\\
		\rowcolor{Gray} $\pi_i$ &  Utility of vendor $i$\\
		$Q$ &  Quality of security patch updates provided by AOSP \\
		\rowcolor{Gray} $\beta$ &  Consumer's security-importance\\
		$T$ &  Consumer's customization-importance\\
		\rowcolor{Gray} $x_j$ &  Consumer's location\\
		$u_j^i$ &  Utility of consumer $j$ for choosing Android type $i$\\
		\rowcolor{Gray} $q^{min}$ &  Minimum level of security from the regulator's point of view\\
		$f_i$ & Fine function for vendor $i$ \\
		\rowcolor{Gray} $F$ &  Monetary value of fine for each unit of violation from $q^{min}$\\
		\hline
	\end{tabular}
\end{table}

\section{Analytical Results}
\label{sec:Ana}
In this section, we analyze our proposed model. Before considering the two stages, we first have to find the market shares of both vendors. To do so, we need to find the point in which a consumer $j$ is indifferent between choosing vendor 1's product and vendor 2's product. This means that a user's preference at this point is identical for the two products. Hence, we have:
\begin{multline}
u_j^1 = u_j^2 \, \, \, \Rightarrow
\beta q_1- p_1 - T\left(x_{j}-a\right)^2 = \beta q_2 - p_2 - T\left(1-b- x_{j}\right)^2.
\label{eq:marketshare1}
\end{multline}

Solving the above equation yields:
%
\begin{equation}
D_1 = 
x_{j} = a + \frac{1 - a- b}{2} + 
\frac{\beta \left(q_1 - q_2\right)}{2T\left(1-a-b\right)} + \frac{p_2 - p_1}{2T\left( 1 - a- b \right)}.
\label{eq:Demand1}
\end{equation}

All of the consumers that are on the left side of $x_{j}$ choose the product of vendor 1. As a result, the market share of vendor 1 is $D_1 = x_j$.
This means that for equal prices and security qualities, vendor 1 controls its own ``turf'' of size $a$ and the consumers located between vendor 1 and vendor 2 that are closer to vendor 1 than vendor 2. The last two terms represent the effect of security quality and price differentiation on the demand, respectively.  

We restrict the model to consumers who definitely choose between these two products, which is a reasonable assumption for a wide range of parameters given the ``cannot-live-without-it'' desirability of modern phones, which is a valid assumption in economics, see~\cite{tirole1988theory}. Hence, the remaining consumers choose vendor 2's product and its demand is:  
\begin{equation}
D_2
=1 - D_1= b + \frac{1 - a- b}{2} + \frac{\beta \left(q_2 - q_1\right)}{2T\left(1-a-b\right)}+\frac{p_1 - p_2}{2T\left( 1 - a- b \right)}.
\label{eq:DemandG}
\end{equation}

If two vendors are at the same location, they provide functionally identical products. For a consumer who takes into account customization, price, and security quality, the factors that matter in this case are security quality and price. To increase their market share, vendors have to decrease their prices or increase their security quality. This will lead to lower product prices and higher costs due to higher security quality, and significantly lower -- and eventually zero -- utility for both vendors. Hence, vendors have no incentives for implementing customizations that result in identical product locations. 


\textbf{Price Competition}: In the following, we state the price Nash equilibrium.
\begin{thm}
The unique price Nash equilibrium always exists, and it is
\begin{align}
p_1^* &= \frac{\beta}{3} \left(q_1 - q_2\right) + T \left(1 - a- b\right) \left(1 + \frac{a -b }{3}\right),
\label{eq:PriceNE1}
\\
p_2^* &= \frac{\beta}{3} \left(q_2 - q_1\right) + T \left(1 - a- b\right) \left(1 + \frac{ b -a }{3}\right).
\label{eq:PriceNEG}
\end{align}
\label{lem:PriceNo}
\end{thm}

Proof of Theorem~\ref{lem:PriceNo} can be found \ifExtendedVersion
in Appendix~\ref{sub:proofPriceNoF}. 
\fi 

In Theorem~\ref{lem:PriceNo}, the price of a product depends on both the security quality and the customization level of both vendors. Further, the price depends on the customization importance $T$ and security-importance $\beta$ constants, which model the consumers in our model. A vendor can increase its price by improving its security quality or customizing its devices more. 


\textbf{Quality and Product Choice:}
To calculate the Nash equilibrium of both vendors in terms of location and security quality, we consider the following optimization problems. 

Vendor 1 maximizes its utility in $q_1$ and $a$ considering that $p_1$ is calculated according to Equation~\ref{eq:PriceNE1}. For vendor 1, we have:

\begin{equation}
\begin{aligned}
& \underset{a, \, q_1}{\text{maximize}}
& & p_1^* D_1 - C_1 \left(a - Z_A\right)^2 - S_1 q_1^2\left( a - Z_A\right)^2 \\
& \text{subject to}
& & p_1^* \geq 0, ~~ 0 \leq a \leq Z_A, ~~ 0 \leq q_1 \leq Q.
\end{aligned}
\label{eq:V1Opt}
\end{equation}

The constraints in the above optimization problem reflect our previous assumptions about the parameters in our model definition. 
For each value of $b$ and $q_2$, the solution of the above optimization problem provides vendor 1's best response. In a similar way, for vendor 2, we have:

\begin{equation}
\begin{aligned}
& \underset{b, \, q_2}{\text{maximize}}
& & p_2^* D_2 - C_2 \left(1-b - Z_A\right)^2 - S_2 q_2^2\left( 1-b - Z_A\right)^2 \\
& \text{subject to}
& & p_2^* \geq 0, ~~ 0 \leq b \leq Z_A, ~~ 0 \leq q_2 \leq Q.
\end{aligned}
\label{eq:V2Opt}
\end{equation}

For given values of $a$ and $q_1$, the above optimization problem provides vendor 2's best response. Based on the Nash equilibrium definition, the intersection of these two optimization problems gives the Nash equilibrium of our proposed game, i.e., $a^*$, $b^*$, $q_1^*$, and $q_2^*$.
In
\ifExtendedVersion
Appendix~\ref{sub:NoteOnOpti}
\fi, we provide our method for solving these two optimization problems and for finding the Nash equilibrium.

\begin{lem}
When consumers take into account security, zero investment in security for both vendors, i.e., $q_1 = q_2 = 0$, is not a Nash equilibrium.
\label{lem:QualInvest}
\end{lem}

Proof of Lemma~\ref{lem:QualInvest} is provided \ifExtendedVersion
in Appendix~\ref{sub:SecInv}. 
\fi

The above lemma shows that when consumers take into account security, then vendors have to invest to improve their security quality. However, it is challenging for the majority of consumers to measure by themselves the security quality of a product, or in this case, to make a comparison between the security quality of many versions of Android provided by the vendors. Consumers mainly rely on information that is made available to them.\footnote{While we have identified a small set of research projects which aim to understand the security impact of customization, e.g., ~\cite{Thomas15,wu2013impact,zhou2014peril}, we are unaware of any well-known market signals regarding the security of different Android versions. The recent FTC initiative to solicit security-relevant data from vendors may contribute to such signals in the future~\cite{FTC}.} However, in the absence of any reliable market signal, any unsubstantiated communication/advertisements by vendors about security quality have to be considered with caution.\footnote{In fact, research by Wu et al. shows that vendors of different reputation (which may also influence perceptions regarding Android security) all suffer from similar challenges due to Android customization \cite{wu2013impact}.}

Previous research has shown that businesses aim to exploit such information barriers. In particular, the theory of \textit{informational market power} posits that when it is hard for consumers to understand and/or observe certain features of a product (e.g., security quality), then businesses are incentivized to underinvest in these product features and rather focus on easily observable aspects such as product design and price~\cite{beales1981efficient}.\footnote{Note that it is not required that businesses have an accurate assessment of the security quality of their own product (or competitors' products) for informational market power to be exploited.} Any effect of informational market power is emphasized by well-known human biases such as \textit{omission neglect} \cite{Kardes07}. This describes the human lack of sensitivity about product features that are not the focus of advertisements or product communications; to paraphrase, consumers will not include in their \textit{perceived utilities} product features which are not emphasized.
Therefore, in Appendix~\ref{app:WithoutSec}, we consider an important baseline case of na\"ive consumers.
In particular, we show that our model is in agreement with what we have seen in practice, i.e., vendors do not invest in security when consumers are na\"ive. Further, we determine under what conditions maximal differentiation, i.e., $a^*=b^*=0$, is Nash equilibrium; see Proposition~\ref{Pro:LOCNEmax} in Appendix~\ref{app:WithoutSec}. 
\vspace{-4mm}

\section{Parameter Selection}
\label{sec:Num}
In the previous section,
the goal of our analyses was to find these six variables: $p_1$, $p_2$, $q_1$, $q_2$, $a$, and $b$. In addition to these six variables, we have six parameters, which are $\beta$, $T$, $C_1$, $C_2$, $S_1$, and $S_2$. In this section, we discuss how we can quantify these six parameters in practice. 
In doing so, we use a reverse approach. First, we measure the values of $p_1$, $p_2$, $q_1$, $q_2$, $a$, and $b$. Then, based on our analyses in the previous section, we calculate the values of the constants in our model. 

\begin{table*}
	\centering
	\tiny
	\begin{tabular}{|c|c|p{1.5cm}|c|c|c|c|c|c|c|c|} \hline
		\multicolumn{5}{|}{} & \multicolumn{2}{|c}{AOSP} & \multicolumn{2}{|c}{vendor} & \multicolumn{2}{|c|}{3rd-party} \\ \hline
		\textbf{Vendor} & \textbf{Device} & \textbf{Version and~Build$\#$} & \textbf{$\#$apps} &  \textbf{$\#$LOC} &  \textbf{$\#$apps} &  \textbf{$\#$LOC} &  \textbf{$\#$apps} &  \textbf{$\#$LOC} & \textbf{$\#$apps} &  \textbf{$\#$LOC} \\ \hline
		HTC (vendor 1) & One X & 4.0.4; CL100532 & 280 & 19M & 29 & 4.7M & 190 & 7.3M & 61 & 7.5M \\ \hline
		Samsung (vendor 2) & Galaxy S3 & 4.0.4; 19300UBALF5 & 185 & 17M & 30 & 6.3M & 119 & 5.6M & 36 & 5.3M \\ \hline
	\end{tabular}
	\caption{Origin of apps in two devices~\cite{wu2013impact}. 
		}
	\label{tab:LOC}
\end{table*} 


\textbf{Location Quantification:} In order to quantify customization and map it to a location, we need to quantify how different two Android versions are in terms of the pre-loaded apps. To do so, we can access the image of an Android OS version, e.g., see~\cite{wu2013impact} and \cite{aafer2016harvesting}, and investigate how many apps a vendor has developed for a specific version. 

To quantify customization, we use the results of Table~\ref{tab:LOC} and calculate the proportion of the code that was developed by a vendor. Note that in our model, we assume that the locations are in the interval $[0,1]$. First, we need to specify the location of $Z_A$ and then select the location of other vendors. 
Here, we assume that $Z_A = 0.5$. In HTC One X (i.e., vendor 1), 7,354,468 LoC were developed by the vendor and 7,550,704 LoC were from third-party apps. This means that about $75.95\%$ of lines of code were added by that vendor to the baseline AOSP version. Here, we interpret this number as the level of difference between its device and AOSP.  In order to keep the value of $a$ in the interval $[0,0.5]$, we let $a = Z_A - (percentage/2)$. Therefore, we have $a = Z_A - (0.7595/2) = 0.1203$. In a similar way, for Samsung Galaxy 3 (i.e., vendor 2) 5,660,569 LOC were developed by the vendor in addition to 5,334,152 LoC selected from third-party apps, which is equal to $63.41\%$ of the total number of LOC. In a similar way, we let $b = 1 - Z_A - (0.6341/2) = 0.1830$.



\textbf{Quality:} To quantify $q_1$ and $q_2$, we use the analysis reported in~\cite{wu2013impact}. According to their analysis, the maximum number of vulnerabilities for a device among these 10 devices is 40. Some of these vulnerabilities are the result of vendor customization. For HTC One X, they have found 15 vulnerabilities, and 10 of these vulnerabilities are due to vendor customization. By dividing the number of vulnerabilities resulting from customization with the maximum number of vulnerabilities, we get $0.25$. In order to calculate security quality, we let $$q_1= 1 - \frac{\# Customization\, Vulnerabilities}{Maximum \# Vulnerabilities} = 0.75.$$ In a similar way, for Samsung Galaxy S3, they found 40 vulnerabilities, and 33 of these are the result of customization. Hence, we have $q_2 =1-\frac{33}{40}= 0.1750$.


\textbf{Price:} The prices of HTC One X and Samsung Galaxy S3 are equal to \euro{170}~\cite{HTC} and \euro{190}~\cite{SamSung}, respectively. GSM Arena (http://www.gsmarena.com/) groups both of these devices as group 4 out of 10 for their price. Here, we consider $p_1 = p_2 = 4$.


\textbf{Parameter Estimation:} By assigning these six parameters into our model analysis, we can calculate the six constants in our model. Here,  we assume that both vendors are completely rational and as result, they have chosen their customization levels, prices, and security qualities according to our proposed model.  Therefore, we can calculate constant parameters in our model in a reverse way. By inserting our quantified parameters, i.e., $a$, $b$, $q_1$, $q_2$, $p_1$, and $p_2$, into Equations~\ref{eq:PriceNE1} and \ref{eq:PriceNEG}, we have two equations and two variables, i.e., $\beta$ and $T$. The answer of this system of equations gives the values of $\beta$ and $T$. Note that these two equations are linear in $T$ and $\beta$. Therefore, the resulting answer is unique.


To calculate the values of $C_1$, $C_2$, $S_1$, and $S_2$, we assume that the measured values of $q_1$, $q_2$, $a$, $b$ form a Nash equilibrium of our game. Since the vendors' strategies are mutual best responses in a Nash equilibrium, $q_1$, $q_2$, $a$, $b$ are solutions to the corresponding best-response equations, i.e., Equations~\ref{eq:QualV1}, \ref{eq:GLoc3}, \ref{eq:QualV2NF}, and \ref{eq:1Loc3}, respectively. We have four variables and four equations. The solution of this system of equations provides the values of $S_1$, $C_1$, $S_2$, $C_2$, which are unique. Therefore, based on the measured values, we have $\beta=0.4362$, $T=5.7414$, $S_1=0.6723$, $C_1=1.4882$, $S_2=4.1338$, and $C_2=2.4875$. 

\vspace{-4mm}


\section{Fine Model and Analysis}
\label{sec:SysModFine}
Our previous analysis shows vendors' poor practice of security issues arising from customization. In particular, we have shown that vendors will not invest in security issues of customization if the consumers do not take security into account. Here, we propose a mechanism to force a vendor to invest in adequate security quality. In doing so, we introduce a regulator whose role is to define the policy that forces every vendor to invest in an adequate level of security. More specifically, we propose a fine function for a regulatory policy which takes as input the vendor's security quality and outputs the monetary value of the fine imposed on the vendor.
In doing so, we define the following fine function for vendor $i$:

\begin{equation}
f_i (q_i) = \begin{cases}
F \left(q^{min} - q_i\right)& \text{if } q^{min} \geq q_i\\
0              & \text{otherwise},
\end{cases}
\label{eq:Fine}
\end{equation}
where $F$ and $q^{min}$ are constants defined by the regulator. $q^{min}$ is the minimum acceptable level of security from the regulator's point of view and the regulator tries to force each vendor to satisfy at least this security level. $F$ is a coefficient relating quality to monetary value and denotes the monetary value of fine for each unit of security violation from $q^{min}$ for a vendor. The monetary value of the fine should be proportional to the market share, since a higher market share of a vendor with security issues results in a higher number of consumers with vulnerabilities. In our model, we multiplied $f_i$ by the market share of that vendor.

In this section, we show that under certain conditions, a regulator can force a vendor to spend on security issues resulting from customization. Moreover, we prove that the product's price \textbf{decreases} as the vendor invests in the adequate level of security imposed by the regulator, for the same value of customization cost. More specifically, our analysis shows that under some conditions, the higher the security quality imposed by the regulator is, the lower the product's price is. 

By imposing a fine, the vendors' utilities change
to the following: 
\begin{align}
\pi_1 = p_1 D_1 - C_1 \left(a- Z_A\right)^2 - S_1 q_1^2\left( a- Z_A\right)^2 - f_1 D_1.
\label{eq:venforUtilF} \\
\pi_2 = p_2 D_2 - C_2 \left(1 - b - Z_A\right)^2 - S_2 q_2^2\left( 1 - b - Z_A\right)^2 - f_2 D_2.
\label{eq:venforUtil2-F}
\end{align}

It is worth mentioning that the consumers' utility does not change. Hence, all of Equations~\ref{eq:marketshare1}, \ref{eq:Demand1}, and \ref{eq:DemandG} are still valid for the case when there is a fine. The validity of these equations implies that the formulae for the vendors' market share is the same for both cases. However, the vendors' equilibrium prices are different compared to the previous case.

Similar to the case without a fine, here we have the same two stages with the same ordering. The regulator's goal is to force the vendors to invest in an adequate security level. Hence, in our analysis, we focus on the case where the regulator forces the vendor to invest in an adequate security quality level. 

Theorem~\ref{thm:PriceNEFine} characterizes both vendors' prices in Nash equilibrium when the regulator imposes a fine.  

\begin{thm}
The Nash equilibrium in prices, which always exists, is
\begin{align}
p_1^* &= \frac{\beta}{3} \left(q_1 - q_2\right) + T \left(1 - a- b\right) \left(1 + \frac{a -b }{3}\right) + \frac{2f_1}{3} + \frac{f_2}{3},
\label{eq:PriceNE1F}
\\
p_2^* &= \frac{\beta}{3} \left(q_2 - q_1\right) + T \left(1 - a- b\right) \left(1 + \frac{ b -a }{3}\right) + \frac{2f_2}{3} + \frac{f_1}{3}.
\label{eq:PriceNEGF}
\end{align} 
\label{thm:PriceNEFine}
\end{thm}
\vspace{-4mm}

This theorem is proved 
\ifExtendedVersion
in Appendix~\ref{sub:proofPriceNE}. 
\fi
By comparing the above two equations with Equations~\ref{eq:PriceNE1} and \ref{eq:PriceNEG}, we observe that the introduction of a fine will increase the product price of the vendors for fixed locations and security level. 
\textbf{Na\"ive Consumers}: Based on Theorem~\ref{thm:PriceNEFine}, by letting $\beta=0$, we can characterize the price NE for na\"ive consumers (which is shown in Appendix~\ref{sub:app-naive}).
Lemma~\ref{lem:QualCond} introduces the sufficient conditions to force vendors to invest in adequate level of security, when consumers do not take into account security.
\begin{lem}
	Both vendors invest in 
    $q_1^*=q_2^*= q^{min}$, if the following conditions are satisfied for the optimal locations of both vendors:
	\begin{align}
	F^2 - 18TS_1 \left(1 - a - b \right)(a - Z_A)^2 \geq 0,
	\label{eq:QualCond1}
	\\
F^2 - 18TS_2 \left(1 - a - b \right)(1- b - Z_A)^2 \geq 0,
	\label{eq:QualCond2}
	\\
3+a-b - \frac{ F q^{min}}{T\left( 1- a - b \right)} \geq 0.
	\label{eq:QualCond3}
	\end{align}
	\label{lem:QualCond}
\end{lem}
\vspace{-4mm}
Proof of the above lemma is provided
\ifExtendedVersion
in Appendix~\ref{sub:qualityNEfine}. 
\fi

Lemma~\ref{lem:Vend1BRFine} calculates the location Nash equilibrium of both vendors considering that the regulator forces the vendors to invest in adequate levels of security.

\begin{lem}
	For given $b$, the vendor 1's best response for location when the consumers do not take into account security and conditions of Lemma~\ref{lem:QualCond} are satisfied, is as follows:
	
	$\bullet \, C_1 + S_1\left(q^{min}\right)^2 \leq \frac{T}{12Z_A}  $: Vendor 1 differentiates its product the most, i.e., $a^*(b)=0$.
	
	$\bullet \, C_1+ S_1\left(q^{min}\right)^2 \geq \frac{T}{9Z_A}$: The positive root of the following quadratic equation is called $a_2$. In this case, for vendor 1 we have $a^*(b) = \min\{a_2,Z_A\}$. 
	
	\begin{multline}
	-3Ta^2 + a \left(2Tb - 10T -36\left(C_1 + S_1\left(q^{min}\right)^2\right)\right) \\ + T\left(b^2-2b-3\right)+36\left(C_1 + S_1\left(q^{min}\right)^2\right)Z_A=0
	\label{eq:qudAFine}
	\end{multline}
	
	$\bullet \, \frac{T}{12Z_A}  < C_1 + S_1\left(q^{min}\right)^2 < \frac{T}{9Z_A} $ and $ b \leq \min\{1 - \sqrt{4 - \frac{36\left(C_1+ S_1\left(q^{min}\right)^2\right)}{T}Z_A},Z_A \} $: Vendor 1 chooses its location as $a^*(b) = \min\{a_2,Z_A\}$.
	
	$\bullet \, \frac{T}{12Z_A} < C_1 + S_1\left(q^{min}\right)^2 < \frac{T}{9Z_A}$ and $1 - \sqrt{4 - \frac{36\left(C_1 + S_1\left(q^{min}\right)^2\right)}{T}Z_A} \leq Z_A$ and $1 - \sqrt{4 - \frac{36\left(C_1 + S_1\left(q^{min}\right)^2\right)}{T}Z_A} \leq b \leq Z_A$: Vendor 1 differentiates its product the most, i.e., $a^*(b)=0$.
	\label{lem:Vend1BRFine}
\end{lem}

	
	
	
	
	
By changing $C_1$ to $C_2$, $S_1$ to $S_2$, $a$ to $b$, and $Z_A$ to $\left(1-Z_A\right)$, in the above lemma, we can derive the same results for vendor 2. 
Proof of Lemma~\ref{lem:Vend1BRFine} is provided
\ifExtendedVersion
in Appendix~\ref{proof:qualFine}. 
\fi

Comparing this lemma with Lemma~\ref{lem:Vend1BR} (Appendix~\ref{sub:NoSec}), we see maximal differentiation occurs when the cost of customization is lower than when there is no fine, since a vendor's cost is affected by both the costs of customization and security quality. Further, according to Equation~\ref{eq:qudAFine}, the location NE depends on $q^{min}$ rather than $F$. However, both F and $q^{min}$ have the effect to satisfy the conditions for forcing a vendor to invest in adequate level of security, i.e., Lemma~\ref{lem:QualCond}.

\vspace{-4mm}

\section{Numerical Illustration}
\label{sec:NumiLL}
In this section, we evaluate our findings numerically. First, we evaluate the case where there are no fines and the consumers are na\"ive.
Second, we evaluate the case in which consumers are na\"ive, but a regulator imposes fines. 
Then, we compare the equilibrium prices and locations in the absence and in the presence of the regulatory fine. Interestingly, we observe that the \textbf{products' prices} (of both vendors) \textbf{decrease} in the presence of fines, and both vendors invest in the adequate level of security $q^{min}$ set by the regulator. Finally, we evaluate the case in which there are no fines but the consumers take into account security quality. 

\begin{figure}[h!]
	\centering
	\begin{subfigure} [$a^*$]{ %
			\includegraphics[width=0.4\textwidth]{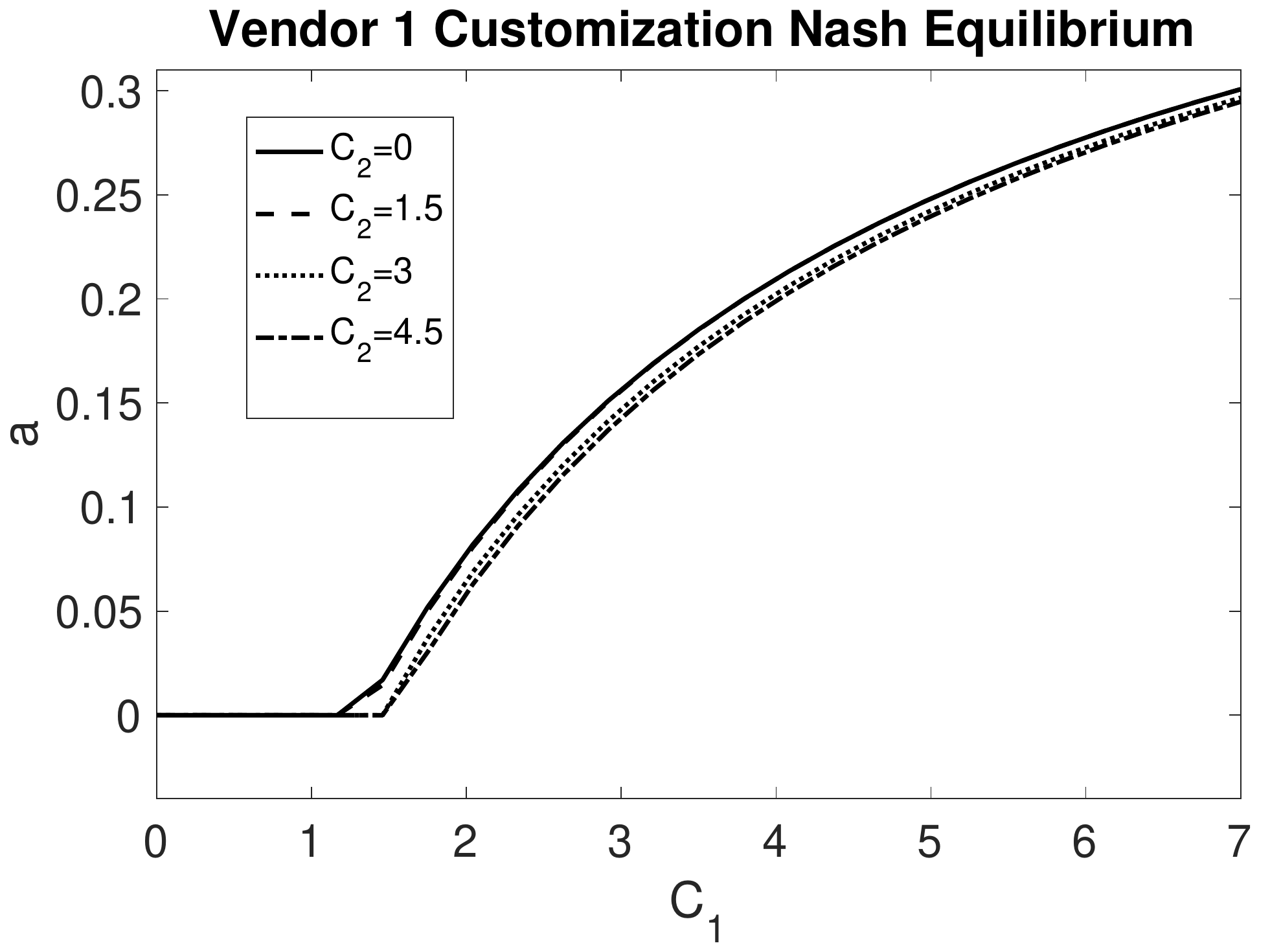}
			\label{fig:a}}
	\end{subfigure}
	\begin{subfigure} [$b^*$]{ %
			\includegraphics[width=0.4\textwidth]{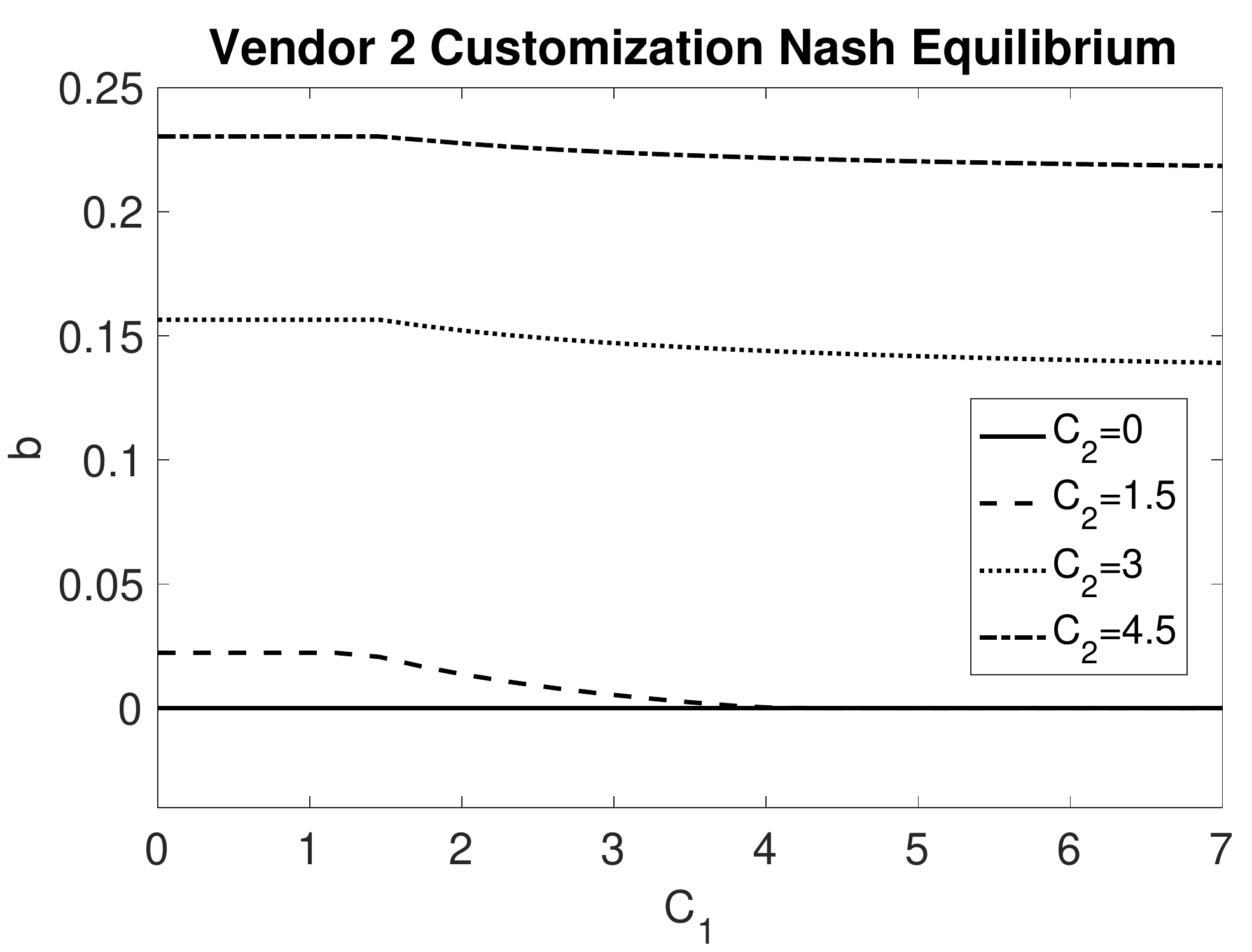}
			\label{fig:b}}
	\end{subfigure}
	\begin{subfigure} [$p_1^*$]{ %
			\includegraphics[width=0.4\textwidth]{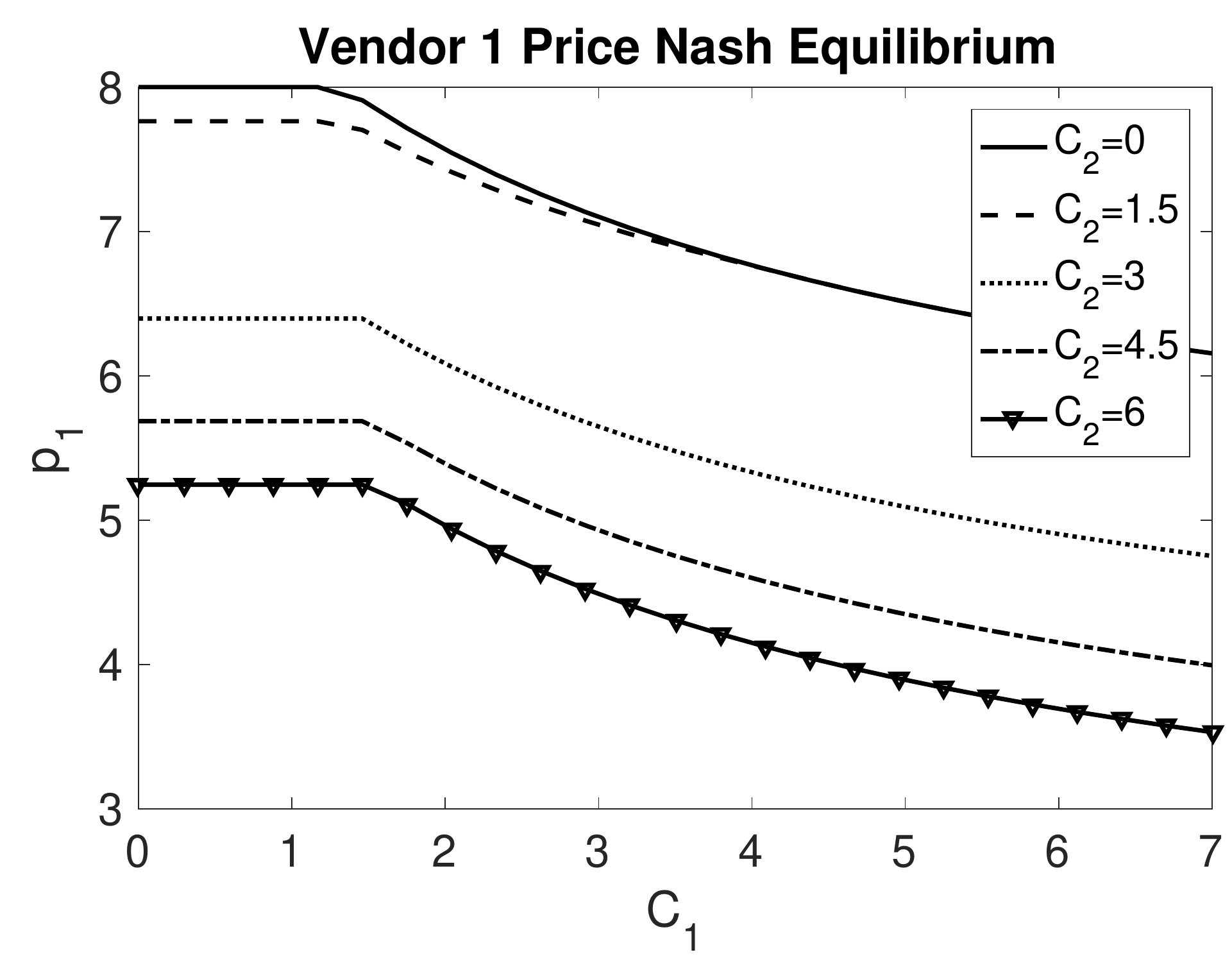}
			\label{fig:p1}}
	\end{subfigure}
	\begin{subfigure} [$p_2^*$]{ %
			\includegraphics[width=0.4\textwidth]{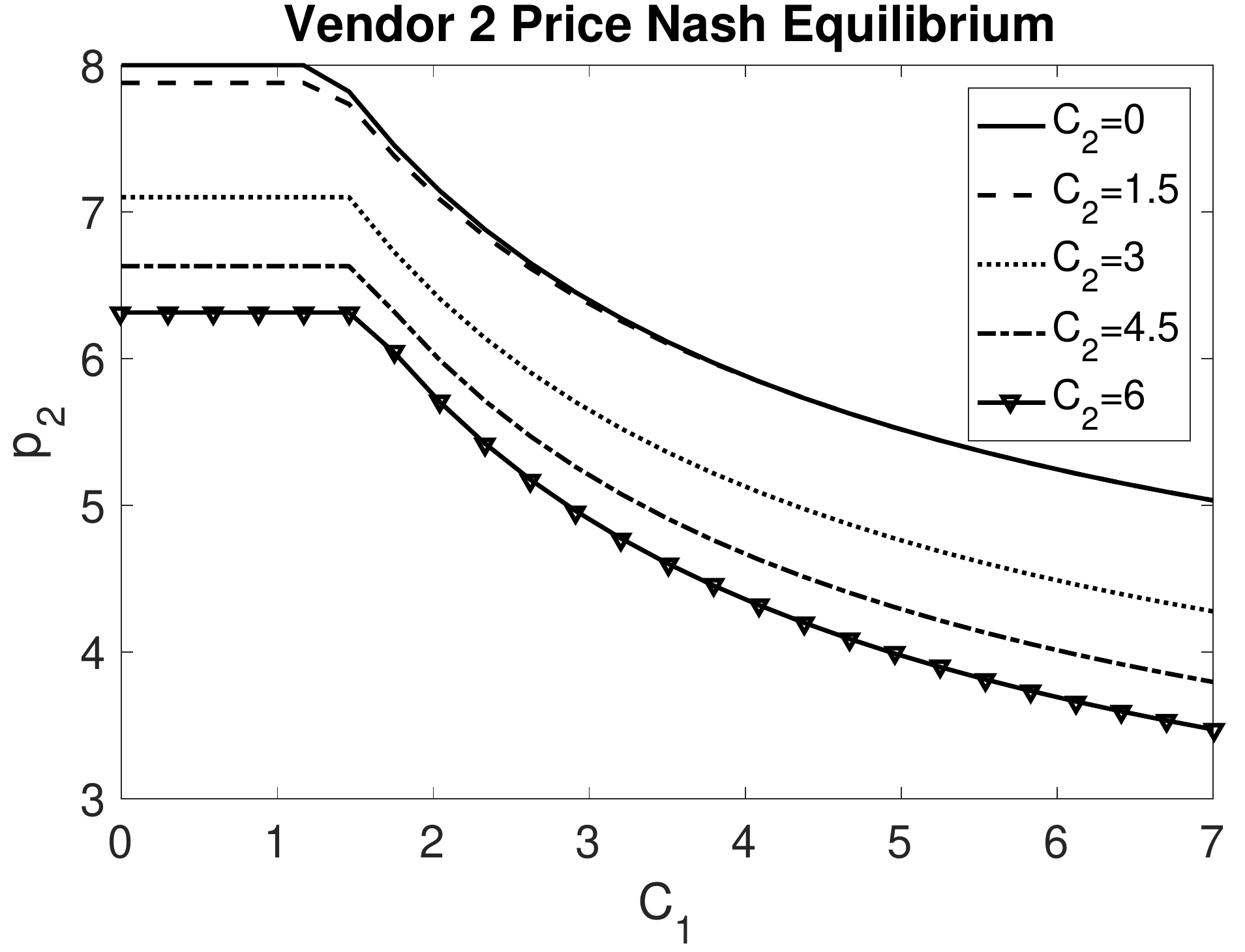}
			\label{fig:p2}}
	\end{subfigure}
	\caption{Equilibrium locations and prices of both vendor 1 and vendor 2 when consumers do not take into account security, there is no fine, and $T=8$.} 
	\label{fig:NoFineNoSec}
\end{figure}

Figure~\ref{fig:NoFineNoSec} shows both vendors' equilibrium locations for various values of customization costs $C_1$ and $C_2$. To find Nash equilibrium, we use Lemma~\ref{lem:Vend1BR} and its counterpart for vendor 2 to calculate each vendor's best-response location. Then, considering that NE is mutual best response, the intersection of these two best responses gives the equilibrium locations of both vendors. Once we have the equilibrium locations, we can easily calculate the equilibrium prices according to Lemma~\ref{lem:PriceNo-Sec}. Based on Figure~\ref{fig:NoFineNoSec}, the higher the cost of customization (e.g., $C_1$) is, the lower a vendor's differentiation from AOSP baseline model is (e.g., the higher the value of $a^*$ is). Note that in Figure~\ref{fig:b} for $C_2=0$, vendor 2 chooses the maximum level of customization (i.e., $b^*=0$) regardless of the vendor 1's customization level. According to Lemma~\ref{lem:Vend1BR} and its counterpart for vendor 2, when the cost of customization is lower than a threshold (i.e., when $C_2 \leq \frac{T}{12\left(1-Z_A\right)}$), vendor 2 chooses the maximum level of customization. As we see in Figure~\ref{fig:p1}, the customization level of vendor 1, i.e., $a^*$, changes only a little with changes in vendor 2's customization cost. For example, when the customization cost of vendor 2 increases, vendor 2 chooses lower level of customization (i.e., higher value of $b^*$), while vendor 1 increases its customization level (i.e., lowers the value of $a^*$) a little compared to its opponent. However, as we observe in Figures~\ref{fig:p1} and \ref{fig:p2}, the prices of both vendors change more significantly compared to their locations with respect to changes in customization costs. Based on Lemma~\ref{lem:PriceNo-Sec} (see Equation~\ref{eq:PriceNE1-No}), a decrease in the customization level of vendor 2 lowers the values of both $(1-a-b)$ and $\left(1+\frac{a-b}{3}\right)$. 
As a result, vendors enter price competition and decrease their prices, which is shown in Figures~\ref{fig:p1} and~\ref{fig:p2}.

\begin{figure}[h!]
	\centering
	\begin{subfigure} [$a^*$]{ %
			\includegraphics[width=0.4\textwidth]{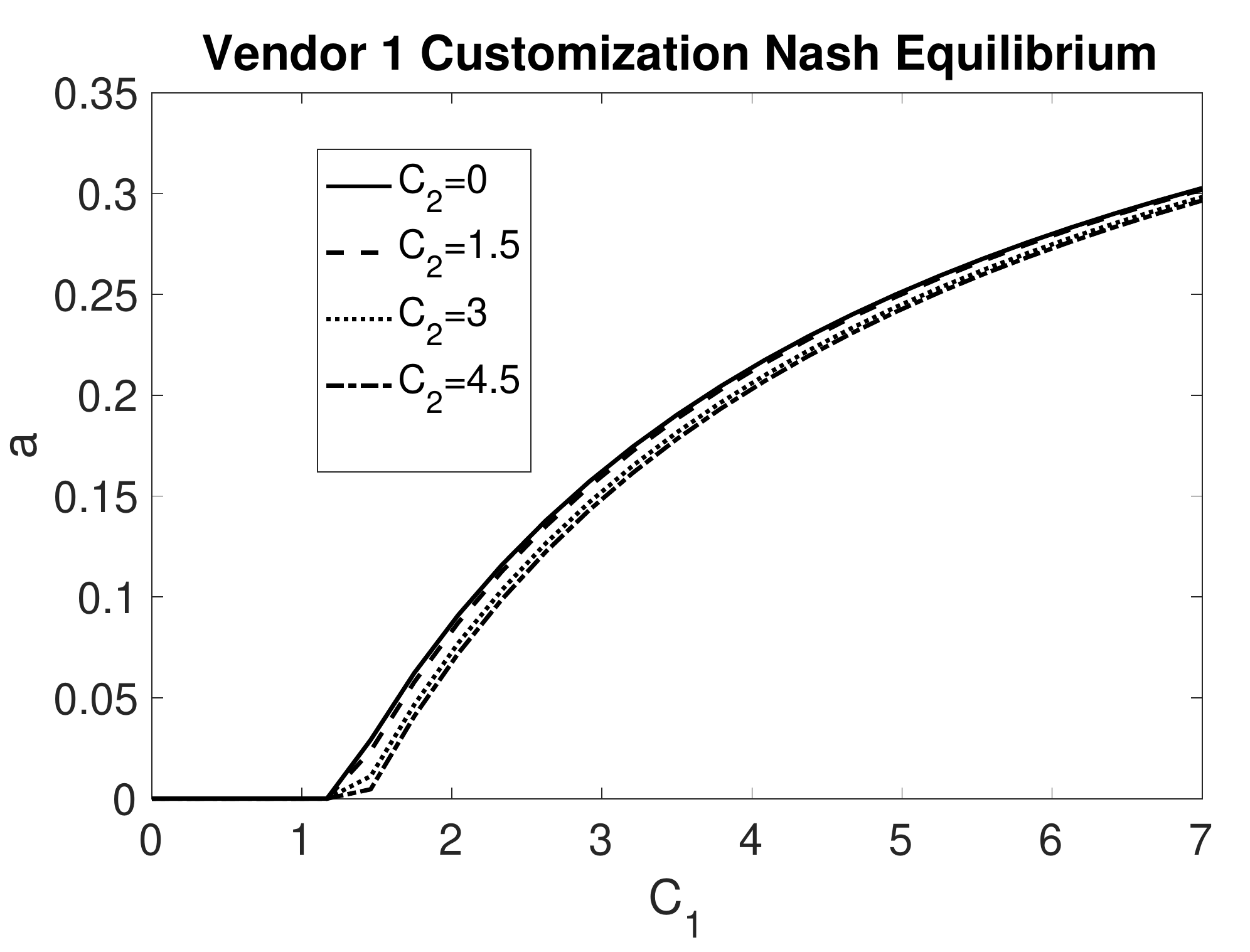}
			\label{fig:aF}}
	\end{subfigure}
	\begin{subfigure} [$b^*$]{ %
			\includegraphics[width=0.4\textwidth]{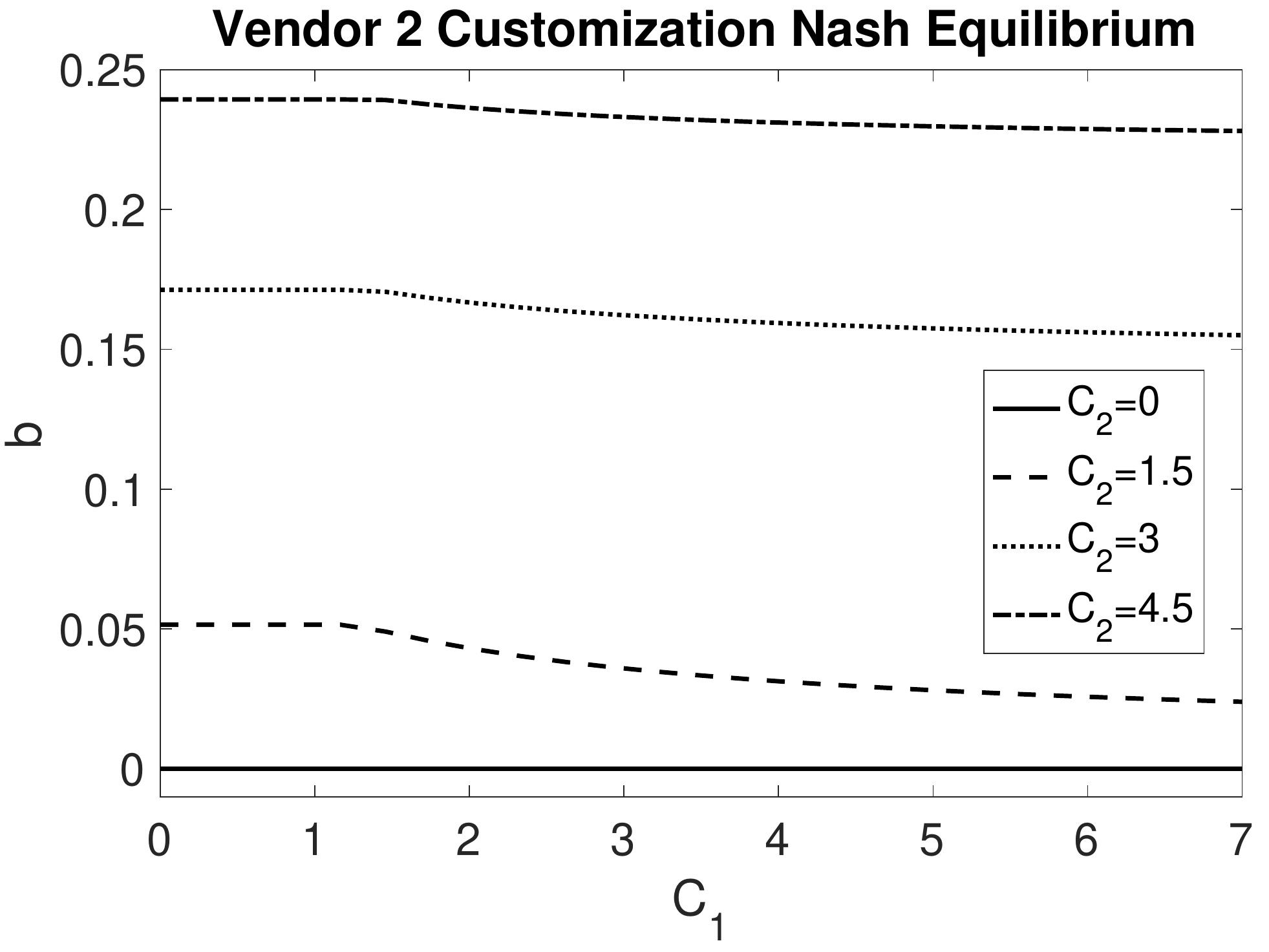}
			\label{fig:bF}}
	\end{subfigure}
	\begin{subfigure} [$p_1^*$]{ %
			\includegraphics[width=0.4\textwidth]{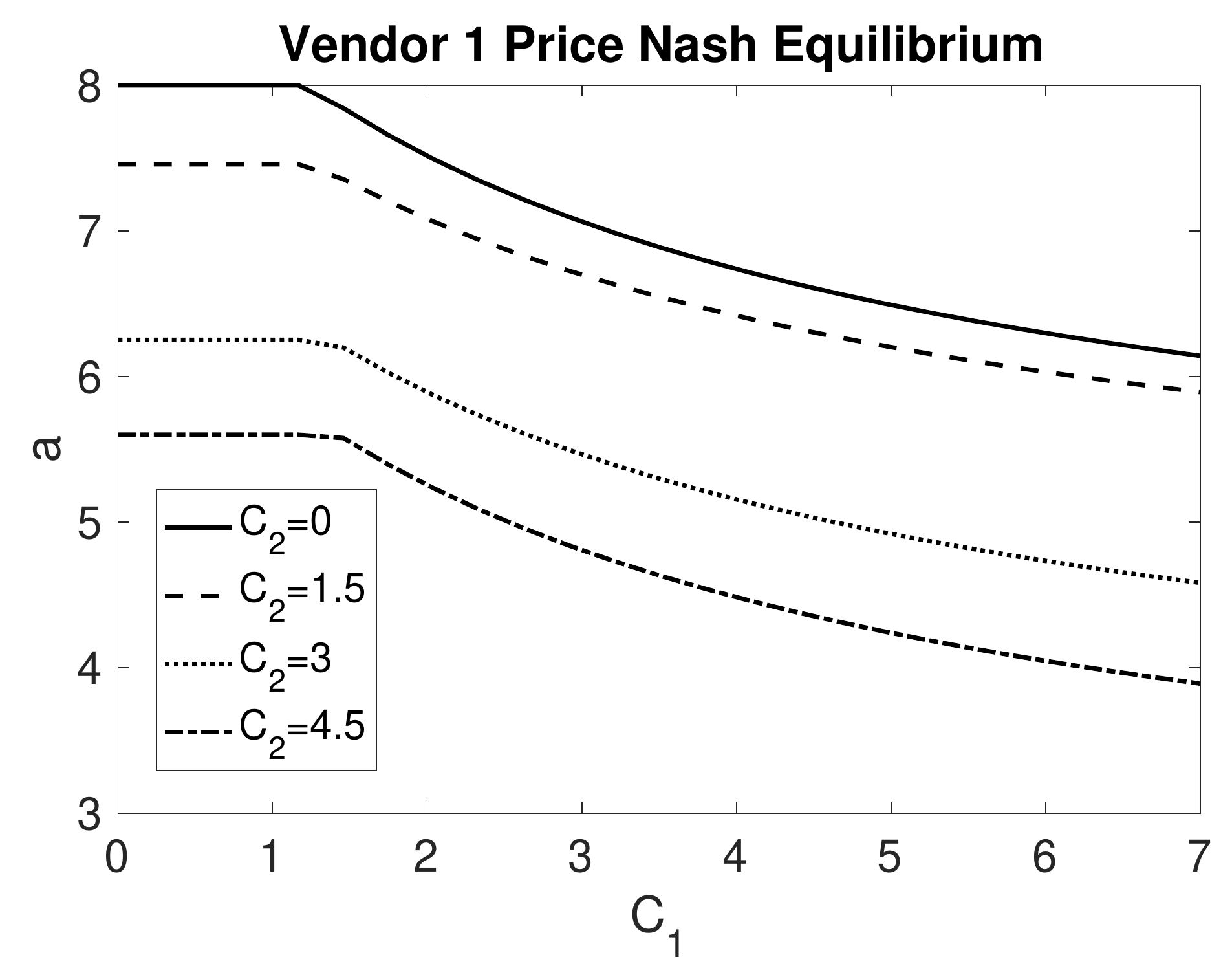}
			\label{fig:p1F}}
	\end{subfigure}
	\begin{subfigure} [$p_2^*$]{ %
			\includegraphics[width=0.4\textwidth]{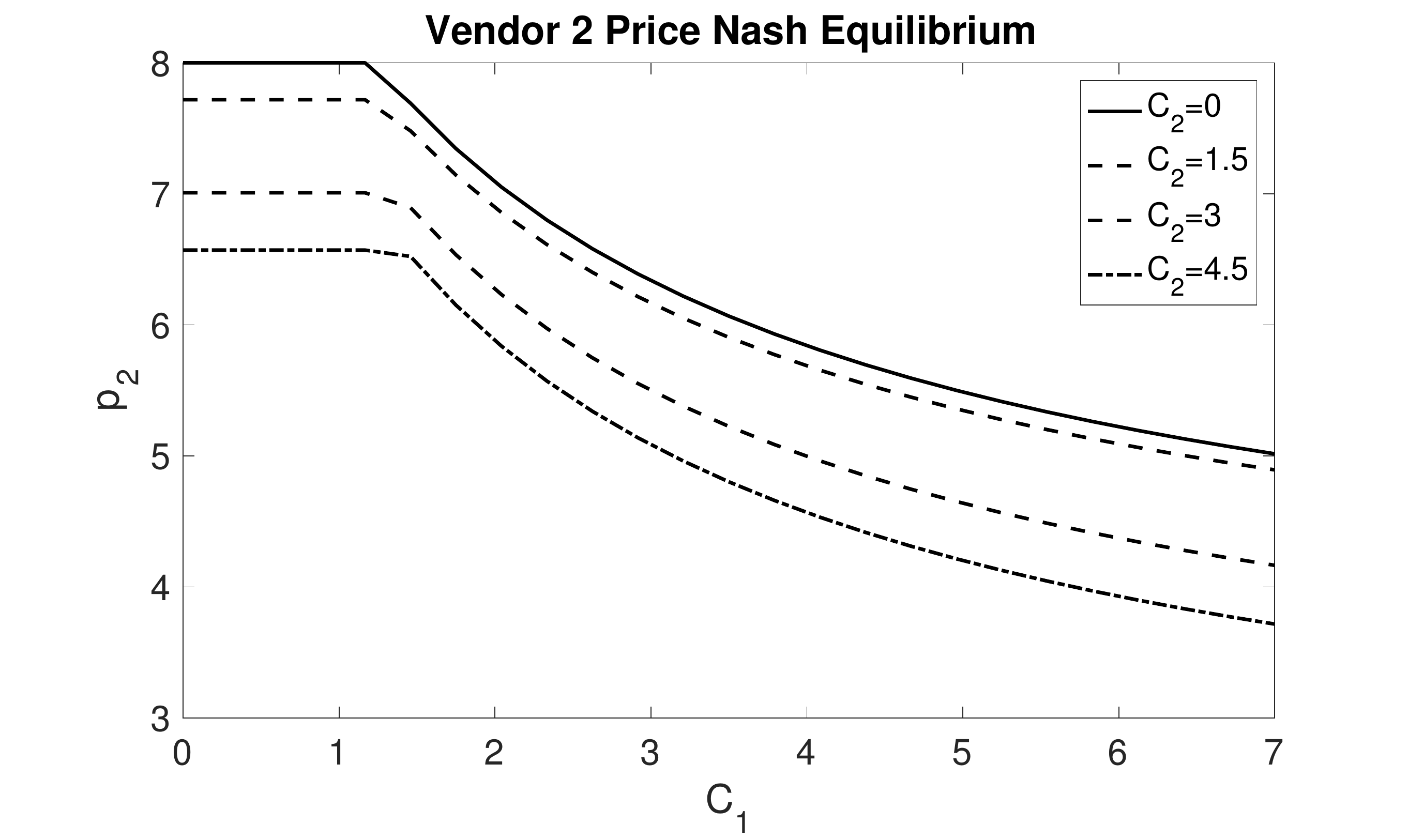}
			\label{fig:p2F}}
	\end{subfigure}

	\caption{Equilibrium locations and prices for various values of $C_1$ and $C_2$. Here, consumers are na\"ive, but there is a regulatory fine, and $T=8$, $S_1 = 0.602$, $S_2=1.54$, $F = 10$, and $q^{min} = 0.4$. For these values of $C_1$ and $C_2$ and choices of $a$ and $b$, both vendors invest in $q^{min}$ set by the regulator.} 
	\label{fig:FineNoSec}
\end{figure}

In Figure~\ref{fig:FineNoSec}, we examine the effect of regulation on location, price, and security quality for various values of $C_1$ and $C_2$. In our evaluation, in the presence of a regulator, the conditions of Lemma~\ref{lem:QualCond} are satisfied. As a result, both vendors invest in $q^{min}$ set by the regulator. 
Similar to the case without any fine, the higher the customization cost (e.g., $C_1$) is, the lower the differentiation from the baseline AOSP is (e.g., the higher the value of $a^*$ is). Similarly, we again observe little changes in a vendor's location in response to changes in its opponent's customization cost and the customization level. Further, the equilibrium prices of both vendors are decreased by an increase in customization costs, since both of them choose lower levels of customization and enter a price competition. Note that even in the presence of fines, vendor 2 chooses the maximum level of customization (i.e., $b^*=0$) when $C_2=0$, considering that its cost of security is proportional to its level of customization and $S_2 > S_1$. The reason for this is that vendor 2 is reluctant to enter a price competition. 
\begin{figure}[h!]
	\centering
	\begin{subfigure} [$a^*$]{ %
			\includegraphics[width=0.4\textwidth]{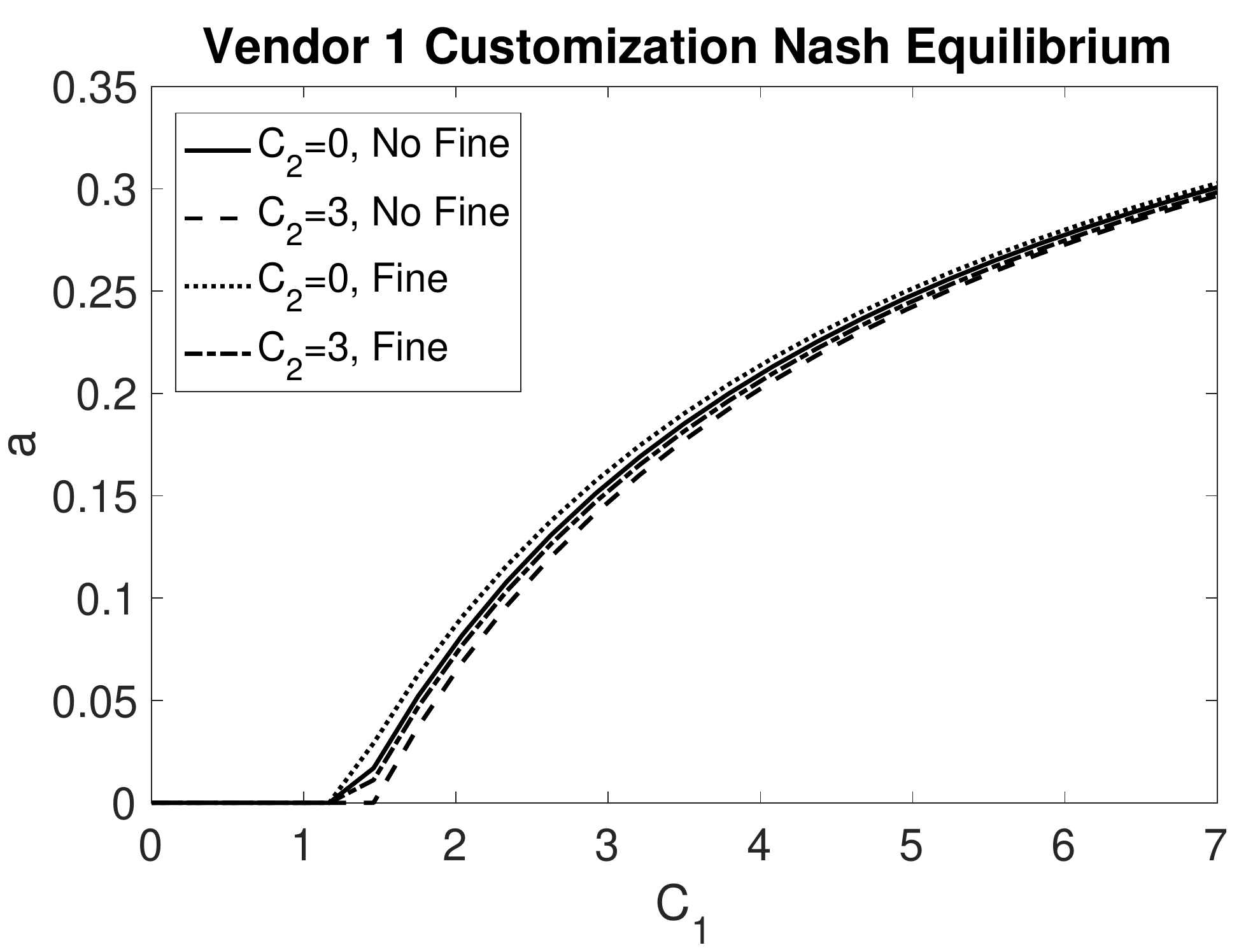}
			\label{fig:aFNF}}
	\end{subfigure}
	\begin{subfigure} [$b^*$]{ %
			\includegraphics[width=0.4\textwidth]{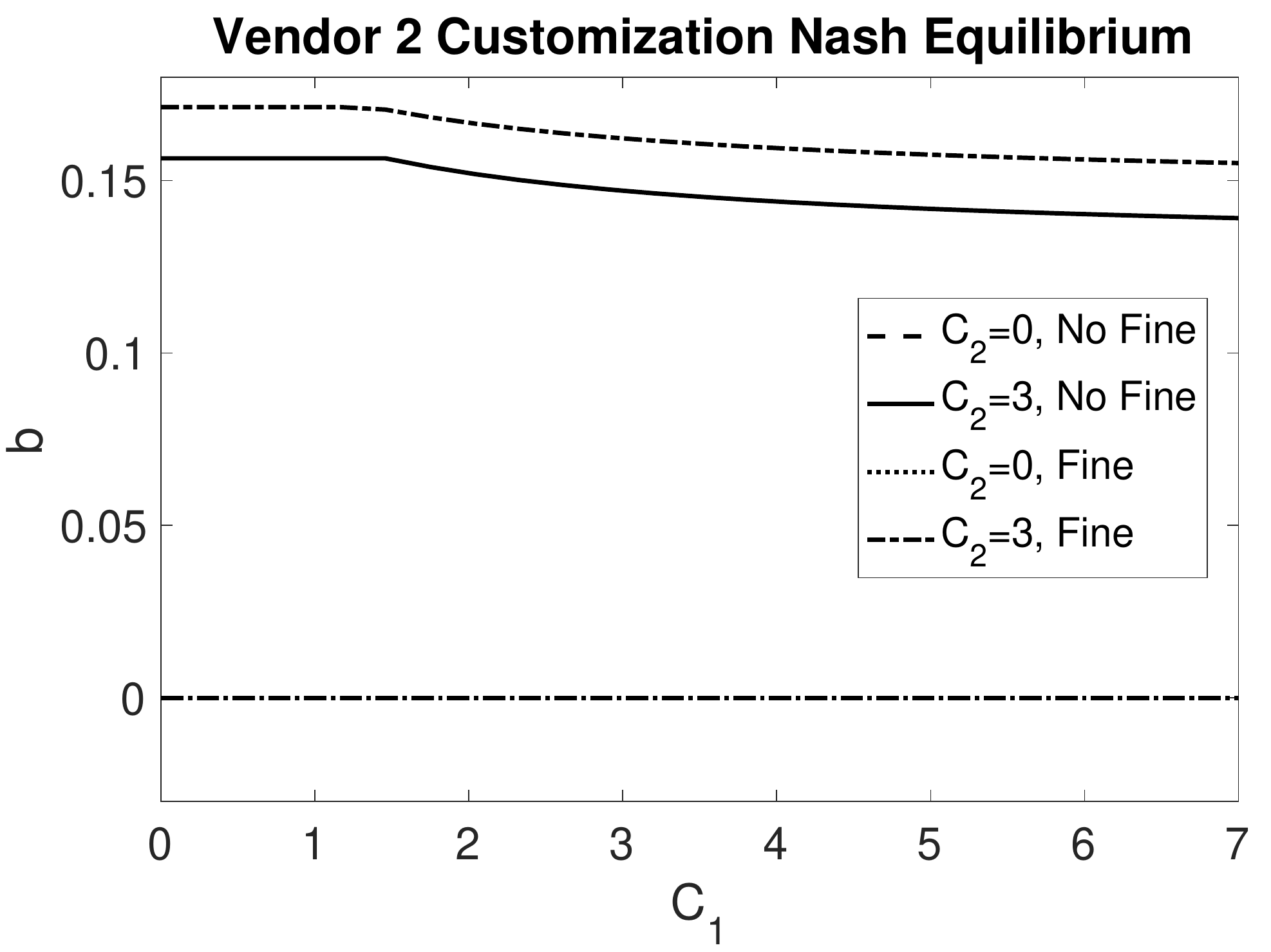}
			\label{fig:bFNF}}
	\end{subfigure}
	\begin{subfigure} [$p_1^*$]{ %
			\includegraphics[width=0.4\textwidth]{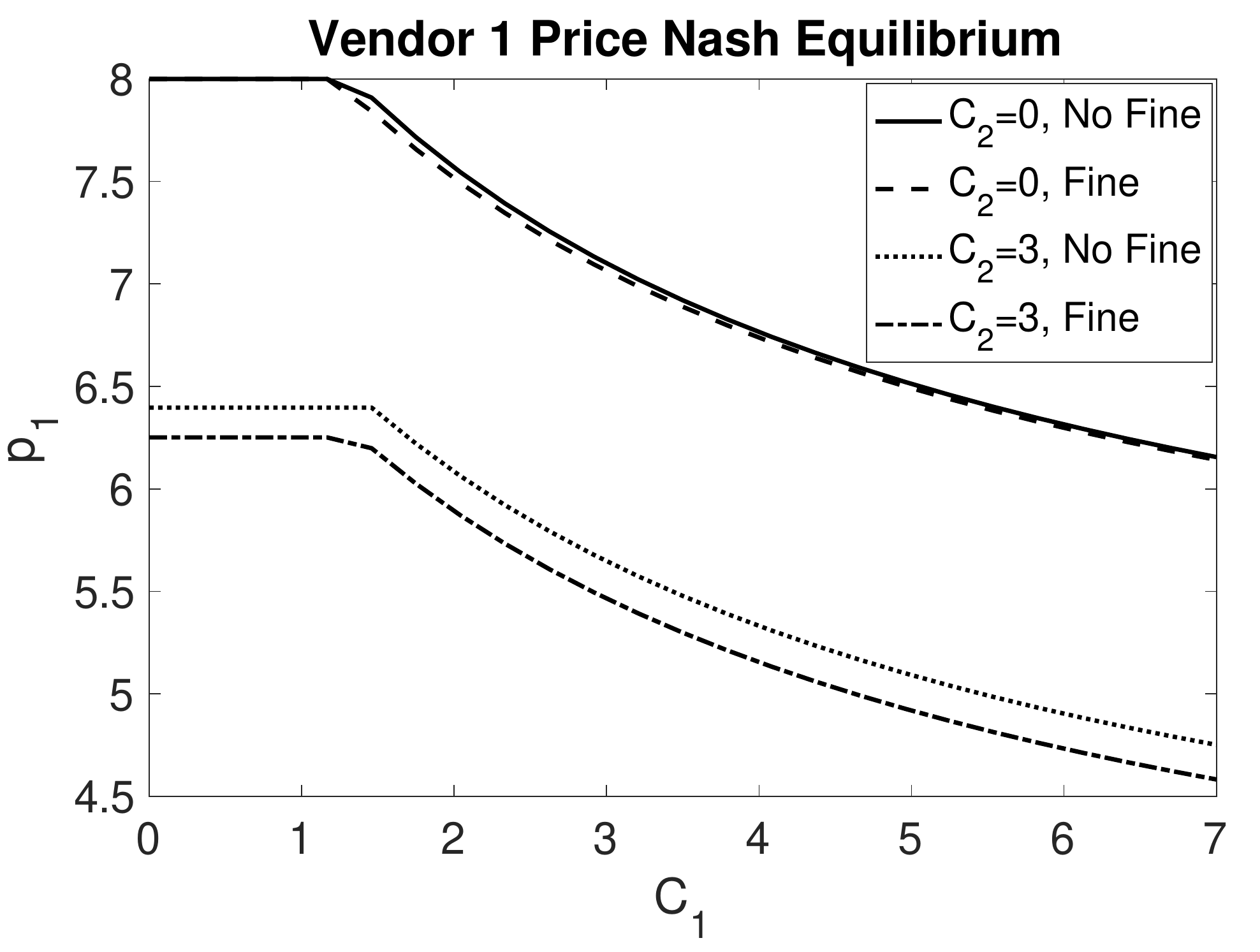}
			\label{fig:p1FNF}}
	\end{subfigure}
	\begin{subfigure} [$p_2^*$]{ %
			\includegraphics[width=0.4\textwidth]{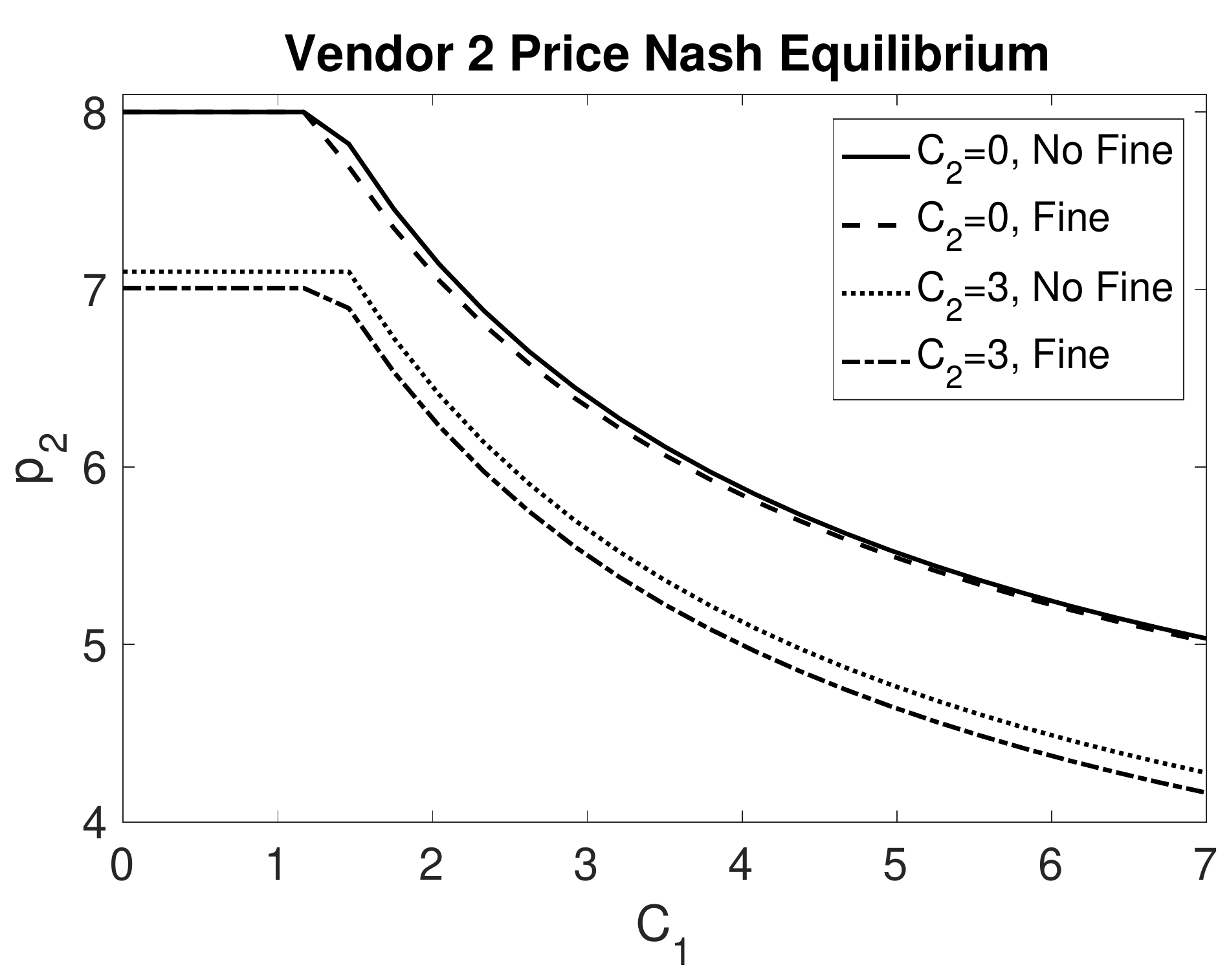}
			\label{fig:p2FNF}}
	\end{subfigure}
    \vspace{-4mm}
	\caption{Comparison between the presence and the absence of a fine for na\"ive consumers. We have $T=8$, $S_1 = 0.602$, $S_2=1.54$, $F = 10$, and $q^{min} = 0.4$.} 
	\label{fig:FineNoSecComp}
\end{figure}

In Figure~\ref{fig:FineNoSecComp}, we compare the equilibria in the presence and the absence of a regulatory fine, when consumers do not take into account security. Based on Figure~\ref{fig:p1FNF}, vendor 1 chooses a lower level of customization when a fine exists. Figure~\ref{fig:p2FNF} shows that vendor 2 chooses the same location in both cases as we discussed earlier when $C_2=0$. For $C_2=3$, vendor 2 chooses a lower level of customization (i.e., higher value of $b^*$) compared to the case when there is no fine. 
Consequently, the prices of both vendors are lower for higher customization costs due to the fact that both vendors are moving closer to the AOSP baseline model. Moreover, the existence of regulation and the fine leads to higher values of $a^*$ and $b^*$ (i.e., lower customization levels) for the same customization costs compared to the case without a fine, since each vendor tries to maximize its utility by avoiding regulatory fine through investing in the minimum level of security quality $q^{min}$. Therefore, each vendor has to pay both the cost of customization as well as the cost of security quality resulting from customization. To decrease these costs, each vendor chooses a lower level of customization. Further, choosing higher values of $a^*$ and $b^*$ (i.e., lower customization level) leads to lower prices for both vendors. Therefore, the existence of a regulatory fine leads to \textbf{more secure products} at \textbf{lower prices} when consumers do not care about security.

To find equilibrium locations and security qualities when consumers take into account security but there is no fine, we calculate each vendor's best-response security quality and location for its opponent's given location and security quality. Then, the Nash equilibrium is the intersection of these best responses. Table~{4} shows the equilibrium for various values of $C_1$, where $T=1.6$, $\beta = 0.6$, $C_2 = 1.3$, $Q=1$, and $S_1 = S_2 = 1$.  
In this case, due to the consumers' security consideration, both vendors invest in security. For $C_1=0$, vendor 1 maximizes its differentiation from baseline AOSP. Because of the consumers' security awareness, vendor 1 invests in security, but at a lower level than $Q$. It is interesting to see that vendor 2 does not differentiate its product from the baseline AOSP version due to maximal differentiation of vendor 1 and consequently, it does not have any security issues resulting from customization (i.e., $q_2^*=Q=1$).

\vspace{2mm}
\begin{table}[h]
	\centering
	\begin{tabular}{|c|c|c|c|c|}\hline
		$C_1$ & $a^*$ & $q_1^*$ & $b^*$ & $q_2^*$ \\ \hline
		$0$ & $0$ & $0.2612$ & $0.5$ & $1$ \\ \hline
		$0.3684$ & $0.2888$ & $1$ & $0.3639$ & $1$  \\ \hline
		$0.7368$ & $0.3195$ & $1$ & $0.3638$ & $1$ \\ \hline
		$1.1053$ & $0.3452$  & $1$ & $0.3637$ & $1$ \\ \hline
		$1.4737$ & $0.3677$ & $1$ & $0.3636$ & $1$ \\ \hline
	\end{tabular}
	\label{tab:NeSecNoF}
	\vspace{2mm}
	\caption{The vendors' equilibrium prices and security qualities for various values of $C_1$. Here, we have $S_1=S_2=1$, $T=1.6$, $\beta=0.6$, $Q=1$, and $C_2=1.3$.}
\end{table}

It is noteworthy that both vendors invest in the maximum level of security when both vendors' customization costs are greater than zero. This observation shows that if all consumers are capable of measuring security quality and it is one of the factors affecting their product choice, then vendors will invest in security. 
Similar to the case where consumers do not take into account security, the higher the customization cost is, the lower the customization level is. In other words, increasing $C_1$ results in higher values of $a^*$. Moreover, changing the value of $C_1$, while $C_2$ is fixed, results in little changes in $b^*$. 


\section{Conclusion}
\label{sec:Conclu}
Our model shows that vendors have to invest in security quality for security-conscious consumers.
Further, for na\"ive consumers, our proposed model captures the fact that vendors underinvest in security. 
To incentivize vendors to invest in security for na\"ive consumers, a regulator may assign a fine to those vendors that do not uphold a desired level of security, which is a well-motivated scenario given Android-related FTC actions \cite{FTC13}. 

We show that the imposed fine structure achieves the expected effect
in addition to changes in the competitive landscape. First, the price of the product decreases for the same cost of customization compared to the case without any fine. Second, a higher level of security quality imposed by the regulator leads to a lower product price, if certain conditions are satisfied. Our findings suggest that requiring higher baseline levels of security investments (as triggered by recent FTC actions \cite{FTC13}) does not impose higher product prices on na\"{\i}ve consumers, which is important from a technology policy perspective. Moreover, increasing consumers' attention about security is substantiated by our analysis as a positive and meaningful factor to address challenges related to informational market power and neglected security efforts. 


\textbf{Acknowledgments:} We thank the anonymous reviewers for their comments. The research activities of Jens Grossklags are supported by the German
Institute for Trust and Safety on the Internet (DIVSI). Aron Laszka's work
was supported in part by the National Science Foundation (CNS-1238959) and
the Air Force Research Laboratory (FA 8750-14-2-0180).

\bibliographystyle{amsplain}
\bibliography{ref}

\appendix
\section{Consumers Without Security Consideration}
\label{app:WithoutSec}
\subsection{Consumers Without Security Consideration}
\label{sub:NoSec}

In this section, we consider a special case of our model, where the consumers do not take into account security. This means that, we have $\beta = 0$ and the derivations to calculate the price and quality Nash Equilibrium follow directly. The following two lemmas state their values.

\begin{Pro}
	The unique price Nash equilibrium, which always exists, is 
	\begin{align}
	p_1^* &=  T \left(1 - a- b\right) \left(1 + \frac{a -b }{3}\right),
	\label{eq:PriceNE1-No}
	\\
	p_2^* &=  T \left(1 - a- b\right) \left(1 + \frac{ b -a }{3}\right).
	\label{eq:PriceNEG-No}
	\end{align}
	\label{lem:PriceNo-Sec}
\end{Pro}

\textbf{Proof}. Proposition~\ref{lem:PriceNo-Sec} can be derived by assigning $\beta = 0$ in Theorem~\ref{lem:PriceNo}. $\qed$

The following corollary represents the vendors' lack of incentives to invest in security-related efforts when consumers cannot measure security quality. 

\begin{Cor}
	In any Nash equilibrium, both vendors invest zero in security, i.e., $q_1^* = q_2^* = 0$.
	\label{lem:Qual-NoSec}
\end{Cor}

\textit{Proof}. Note that when $\beta=0$, the market share of each vendor does not depend on security quality. In order to calculate $q_1$, we have:

\begin{equation}
\frac{d \pi_1}{d q_1} = -2q_1S_1 \left(a - Z_A\right)^2,
\label{eq:QualV1-NoSec}
\end{equation}

which is decreasing for non-negative values of $q_1$. Therefore, we have $q_1^* = 0$. In a similar way, for vendor 2 we have $q_2^*=0$. $\qed$

The following lemma represents vendor 1's level of customization best response, i.e., $a$, given the customization level of its opponent, i.e., $b$.

\begin{lem}
For given $b$, vendor 1's best response for location when the consumers do not take security into account is as follows:
	
$\bullet \, C_1 \leq \frac{T}{12Z_A}$: Vendor 1 differentiates its product as much as possible, i.e., $a^*(b)=0$.

$\bullet \, C_1 \geq \frac{T}{9Z_A}$: The positive root of the following quadratic equation in $a$ is called $a_2$. In this case, for vendor 1 we have $a^*(b) = \min\{a_2,Z_A\}$. 

\begin{multline}
-3Ta^2 + a \left(2Tb - 10T -36C_1\right) + T\left(b^2-2b-3\right)+36C_1Z_A=0
\label{eq:qudA}
\end{multline}

$\bullet \, \frac{T}{12Z_A} < C_1 < \frac{T}{9Z_A}$ and $0 \leq b \leq \min\{Z_A, 1 - \sqrt{4 - \frac{36C_1}{T}Z_A}\}$: Vendor 1 chooses its location as $a^*(b) = \min\{a_2,Z_A\}$.

$\bullet \, \frac{T}{12Z_A} < C_1 < \frac{T}{9Z_A}$ and $1 - \sqrt{4 - \frac{36C_1}{T}Z_A} \leq Z_A$ and $1 - \sqrt{4 - \frac{36C_1}{T}Z_A} \leq b \leq Z_A$: Vendor 1 differentiates its product the most, i.e., $a^*(b)=0$.
\label{lem:Vend1BR}
\end{lem}

The above lemma represents that if the cost of customization is lower than a threshold, i.e., $C_1 \leq \frac{T}{12Z_A}$, vendor 1 will differentiate its product the most regardless of its opponent's level of customization, i.e., $a^*=0$. The reason is that when the cost of customization is low, the vendor chooses a location in order to increase the price, see Equation~\ref{eq:PriceNE1-No}, and to prevent price competition. 
By changing $C_1$ to $C_2$, $S_1$ to $S_2$, $a$ to $b$, and $Z_A$ to $\left(1-Z_A\right)$, in the above lemma, we can derive the same results for vendor 2.

	
	
	
	
	

Proof of Lemma~\ref{lem:Vend1BR} 
is provided
\ifExtendedVersion
in Appendix~\ref{sub:proofLocNo}. 
\fi

\begin{Pro}
When consumers do not take security into account, $C_1 \leq \frac{T}{12Z_A}$, and $C_2 \leq \frac{T}{12(1-Z_A)}$ location Nash equilibrium is $a^*=b^* = 0$, i.e., maximal differentiation.
\label{Pro:LOCNEmax}
\end{Pro}

\textit{Proof}. The Nash equilibrium is the mutually best response. Based on Lemma~\ref{lem:Vend1BR} and its corresponding vendor's 2 best response, 
for these cost of customizations, both vendors' best responses are equal to zero. $\qed$

For other values of $C_1$ and $C_2$, we calculate Nash equilibria numerically. 

\subsection{Price Equilibrium for Na\"ive Consumers in the Presence of Fine}
\label{sub:app-naive}

\begin{Cor}
	The Nash equilibrium in prices, which always exists, is
	\begin{align}
	p_1^* &=  T \left(1 - a- b\right) \left(1 + \frac{a -b }{3}\right) + \frac{2f_1}{3} + \frac{f_2}{3},
	\label{eq:PriceNE1FNOs}
	\\
	p_2^* &= T \left(1 - a- b\right) \left(1 + \frac{ b -a }{3}\right) + \frac{2f_2}{3} + \frac{f_1}{3}.
	\label{eq:PriceNEGFNos}
	\end{align} 
	\label{thm:PriceNEFineNos}
\end{Cor}
\section{Proof}
\label{app:proof}
\subsection{Price Nash Equilibrium without Fine}
\label{sub:proofPriceNoF}

First, we take the partial derivative with respect to price for vendor 1 and have:

\begin{multline}
\frac{\partial \pi_1}{\partial p_1} = D_1 \left(a,b,p_1,p_2\right) + p_1  \frac{\partial D_1 \left(a,b,p_1,p_2\right)}{\partial p_1}\\
=\frac{p_2 - p_1}{2T\left(1-a-b\right)} + a +  \frac{1 -a -b}{2} 
+ \frac{\beta \left(q_1 - q_2\right)}{2T\left(1 -a - b\right)} + p_1 \left(\frac{-1}{2T\left(1 -a - b\right)} \right),
\label{eq:PricePartial1}
\end{multline}

the above equation is linear with respect to $p_1$ and the coefficient of $p_1$ is negative. To find the utility-maximizing price, we set the above value equal to zero, which gives us the following:
\begin{multline}
\frac{\partial \pi_1}{\partial p_1} = 0 \, \, \, \Rightarrow \, \, \, 
p^{crit}_1 =  \frac{p_2}{2} +   \frac{T}{2} \left(1 -a - b \right)\left(1 + a -b \right) + \frac{\beta \left(q_1 - q_2\right)}{2}.
\label{eq:PricePartial1-2}
\end{multline} 

Note that $\pi_1$ is increasing in $[0, p_1^{crit}]$ and decreasing in $[p_1^{crit}, \infty]$, since the coefficient of $p_1$ is negative in Equation~\ref{eq:PricePartial1}. Therefore, vendor 1's utility is maximized at $p_1^* = p^{crit}_1$. The value of $p^*_1$ depends on its location, its rival location, security quality differences of two vendors, and vendor 2's product price.  

In a similar way, for vendor 2 we have:

\begin{multline}
\frac{\partial \pi_2}{\partial p_2} =D_2 \left(a,b,p_1,p_2\right) + p_2  \frac{\partial D_2 \left(a,b,p_1,p_2\right)}{\partial p_2}
= \frac{p_1 - p_2}{2T\left(1-a-b\right)} + b + \frac{1 -a -b}{2} + \\
\frac{\beta \left(q_2 - q_1\right)}{2T\left(1-a-b\right)} + p_2 \left(\frac{-1}{2T\left(1 -a - b\right)} \right).
\label{eq:PricePartialG}
\end{multline}

This is linear in $p_2$ and its coefficient is negative. The critical point is:
\begin{multline}
\frac{\partial \pi_2}{\partial p_2} = 0 \, \, \, \Rightarrow \, \, \, 
 p^{crit}_2 =  \frac{p_1}{2} +   \frac{T}{2} \left(1 -a - b \right)\left(1 - a +b \right) + \frac{\beta \left(q_2 - q_1\right)}{2T}.
\label{eq:PricePartialG-2}
\end{multline}

It is straightforward to see that $\pi_2$ is increasing in $[0, p_2^{crit}]$ and decreasing in $[p_2^{crit}, \infty]$. Therefore, $p_2^{crit}$ is local maximum. 

By combining Equations~\ref{eq:PricePartial1-2} and \ref{eq:PricePartialG-2}, we can derive the Nash equilibrium in prices which always exists. $\qed$

\subsection{Optimization Problem with No Fine}
\label{sub:NoteOnOpti}

First, we calculate best-response $q_1$ by taking the derivative of Equation~\ref{eq:venforUtil1} with respect to it and we have:
\begin{equation}
\frac{d \pi_1}{d q_1} =  \frac{\partial \pi_1}{\partial q_1} + \frac{\partial \pi_1}{\partial p_1} \frac{\partial p_1}{\partial q_1} + \frac{\partial \pi_1}{\partial p_2} \frac{\partial p_2}{\partial q_1} .
\label{eq:QualPartial1}
\end{equation}

Note that in the above equation, we are going to maximize $\pi_1$ with respect to $q_1$ given that Equation~\ref{eq:PriceNE1} is satisfied. According to the envelope theorem, vendor 1 maximizes its utility with respect to the price in the second stage, where $\frac{\partial \pi_1}{\partial p_1} = 0$. Therefore, we have:

\begin{multline}
\frac{d \pi_1}{d q_1} =  \frac{\partial \pi_1}{\partial q_1} + \frac{\partial \pi_1}{\partial p_2} \frac{\partial p_2}{\partial q_1} \, \, \, \Rightarrow \\
\frac{d \pi_1}{d q_1}  =  -2S_1 q_1 \left(a - Z_A\right)^2 + p^*_1 \frac{\partial D_1}{\partial q_1} + p_1^* \frac{\partial D_1}{\partial p_2} \frac{\partial p_2^*}{\partial q_1}.
\label{eq:QualPartial1-1}
\end{multline}

By using Equations~\ref{eq:venforUtil1}, \ref{eq:PriceNE1}, and \ref{eq:PriceNEG}, we have:

\begin{multline}
\frac{d \pi_1}{d q_1} = -2S_1 q_1 \left(a - Z_A\right)^2 
+ p^*_1 \left(\frac{\beta}{2T\left(1-a-b\right)}\right) - p_1^* \left(\frac{\beta}{6T\left(1-a-b\right)}\right) \\
= -2S_1 q_1 \left(a - Z_A\right)^2 + p^*_1 \left(\frac{\beta}{3T\left(1-a-b\right)}\right).
\end{multline}

According to Equation~\ref{eq:PriceNE1}, $p_1^*$ is also function of $q_1$. By incorporating Equation~\ref{eq:PriceNE1} into the above equation, we have:

\begin{multline}
\frac{d \pi_1}{d q_1} = q_1 \left(-2S_1 \left(a - Z_A\right)^2 + \frac{\beta^2}{9T\left(1-a-b\right)}\right) 
+ \frac{\beta}{9} \left( 3+a-b - \frac{\beta q_2}{T\left(1-a-b\right)} \right).
\label{eq:QualV1}
\end{multline}

The above equation is linear in terms of $q_1$. To make our analysis simpler, we define the following parameters:

\begin{multline}
A_1 = -2S_1 \left(a - Z_A\right)^2 + \frac{\beta^2}{9T\left(1-a-b\right)}, 
B_1 = \frac{\beta}{9} \left( 3+a-b - \frac{\beta q_2}{T\left(1-a-b\right)} \right).
\end{multline}

According to the sign of these two parameters, we have:

\begin{equation}
opt_{q_1} \left(a,b,q_2\right) =
\begin{cases}
Q & \text{if $A_1 \geq 0 \, , B_1 \geq 0$} \\
\min\{\frac{-B_1}{A_1}, Q\} & \text{if $A_1 < 0 \, , B_1 \geq 0$} \\
\argmax_{q_1 \in \{\bar{q}_1,Q\}} \pi_1 & \text{if $A_1 \geq 0 \, , B_1 < 0$} \\
\bar{q}_1 & \text{if $A_1 < 0 \, , B_1 < 0$}. \\
\end{cases}
\label{eq:BRq1NoF}
\end{equation}

Where $\bar{q}_1 = q_2 - \frac{T}{\beta} \left(1-a-b\right)\left(3+a-b\right)$. The reasons behind the above formula are as follows:

\begin{itemize}
	\item $A_1 \geq 0 \, , B_1 \geq 0$: In this case, $\pi_1$ is increasing for any values $q_1 \geq 0$. Because of our constraint, $\pi_1$ is maximized at $Q$. 
	\item $A_1 < 0 \, , B_1 \geq 0$: $\pi_1$ is increasing in $[0,\frac{-B_1}{A1}]$ and decreasing in $[\frac{-B_1}{A_1}, \infty]$.
	\item $A_1 \geq 0 \, , B_1 < 0$: $\pi_1$ is decreasing in $[0,\frac{-B_1}{A1}]$ and increasing in $[\frac{-B_1}{A_1}, \infty]$. In this case, two candidate points are $0$ and $Q$. Note that when $B_1 < 0$, $q_1=0$ gives negative $p_1$ which is not acceptable. Hence, vendor 1 chooses the lowest value of $q_1$ that gives non-negative $p_1$. This consideration gives $\bar{q}_1$ as one the candidate points.
	\item $A_1 < 0 \, , B_1 < 0$: $\pi_1$ is decreasing for all non-negative values of $q_1$, but $q_1=0$ gives negative $p_1$. Similar to the previous case, $\pi_1$ is maximized at $\bar{q}_1$. 
\end{itemize}



In a similar way for vendor 2 we have:

\begin{equation}
\frac{d \pi_2}{d q_2} =  \frac{\partial \pi_2}{\partial q_2} + \frac{\partial \pi_2}{\partial p_2} \frac{\partial p_2}{\partial q_2} + \frac{\partial \pi_2}{\partial p_1} \frac{\partial p_1}{\partial q_2} .
\label{eq:QualPartial2}
\end{equation}

According to the envelope theorem, vendor 2 maximizes its utility with respect to the price in the second stage where $\frac{\partial \pi_2}{\partial p_2} = 0$. Therefore, we have:

\begin{multline}
\frac{d \pi_2}{d q_2} =  \frac{\partial \pi_2}{\partial q_2} + \frac{\partial \pi_2}{\partial p_1} \frac{\partial p_1}{\partial q_2} \, \, \, \Rightarrow \\
\frac{d \pi_2}{d q_2}  =  -2S_2 q_2 \left(1 - b - Z_A\right)^2 + p^*_2 \frac{\partial D_2}{\partial q_2} + p_2^* \frac{\partial D_2}{\partial p_1} \frac{\partial p_1^*}{\partial q_2}.
\label{eq:QualPartial1-2}
\end{multline}

By using Equations~\ref{eq:venforUtil2}, \ref{eq:PriceNE1}, and \ref{eq:PriceNEG}, we have:

\begin{multline}
\frac{d \pi_2}{d q_2} = -2S_2 q_2 \left(1-b - Z_A\right)^2 
+ p^*_2 \left(\frac{\beta}{2T\left(1-a-b\right)}\right) - p_2^* \left(\frac{\beta}{6T\left(1-a-b\right)}\right) \\
= -2S_2 q_2 \left(1-b - Z_A\right)^2 + p^*_2 \left(\frac{\beta}{3T\left(1-a-b\right)}\right).
\end{multline}

According to Equation~\ref{eq:PriceNEG}, $p_2^*$ is also function of $q_2$. By incorporating Equation~\ref{eq:PriceNEG} into the above equation, we have:

\begin{multline}
\frac{d \pi_2}{d q_2} = q_2 \left(-2S_2 \left(1-b - Z_A\right)^2 + \frac{\beta^2}{9T\left(1-a-b\right)}\right) \\ 
+ \frac{\beta}{9} \left( 3 - a + b - \frac{\beta q_1}{T\left(1-a-b\right)} \right).
\label{eq:QualV2NF}
\end{multline}

The above equation is linear in terms of $q_2$. To make our analysis simpler, we define the following parameters:

\begin{multline}
A_2 = -2S_2 \left(1 - b - Z_A\right)^2 + \frac{\beta^2}{9T\left(1-a-b\right)}, \\
B_2 = \frac{\beta}{9} \left( 3 - a + b - \frac{\beta q_1}{T\left(1-a-b\right)} \right).
\end{multline}

According to the sign of these two parameters, we have:

\begin{equation}
opt_{q_2} \left(a,b,q_1\right) =
\begin{cases}
Q & \text{if $A_2 \geq 0 \, , B_2 \geq 0$} \\
\min\{\frac{-B_2}{A_2}, Q\} & \text{if $A_2 < 0 \, , B_2 \geq 0$} \\
\argmax_{q_2 \in \{\bar{q}_2,Q\}} \pi_2 & \text{if $A_2 \geq 0 \, , B_2 < 0$} \\
\bar{q}_2 & \text{if $A_2 < 0 \, , B_2 < 0$}. \\
\end{cases}
\label{eq:BRq2NoF}
\end{equation}

Where $\bar{q}_2 = q_1 - \frac{T}{\beta} \left(1-a-b\right)\left(3-a+b\right)$.

In order to calculate the optimal location of each player, we take the derivative with respect to location. First, we start with vendor 1 and we have:
\begin{equation}
\frac{d \pi_1 }{d a} = \frac{\partial \pi_1 }{\partial p_1} \frac{\partial p_1}{\partial a} + \frac{\partial \pi_1 }{\partial a} + \frac{\partial \pi_1}{\partial p_2} \frac{\partial p_2}{\partial a}.
\label{eq:GLoc1}
\end{equation}

Note that in the above equation, we are going to maximize $\pi_1$ given that Equation~\ref{eq:PriceNE1} is satisfied. According to the envelope theorem, vendor 1 maximizes its utility with respect to the price in the second period where $\frac{\partial \pi_1}{\partial p_1} = 0$. Therefore, we have:
\begin{multline}
\frac{\partial \pi_1 }{\partial a} = \frac{\partial \pi_1 }{\partial p_2} \frac{\partial p_2^*}{\partial a} + \frac{\partial \pi_1 }{\partial a} = - 2C_1 \left(a - Z_A\right) \\
- 2S_1 q_1^2 \left(a - Z_A\right) + p_1^* \left( \frac{\partial D_1}{\partial a}  + \frac{\partial D_1}{\partial p_2} \frac{ \partial p^*_2}{\partial a}\right) .
\label{eq:GLoc2}
\end{multline}

By using Equations~\ref{eq:Demand1}, \ref{eq:PriceNE1}, and \ref{eq:PriceNEG}, we get:
\begin{multline}
\frac{\partial D_1}{\partial a} = \frac{1}{2} + \frac{ \beta \left(q_1 - q_2 \right) + p_1^* - p_2^*}{2T\left(1 - a - b\right)^2} = 
\frac{3 - 5a - b}{6\left(1 - a - b\right)} + \frac{\beta \left(q_1 - q_2\right)}{6T \left(1-a-b\right)^2}.
\label{eq:DemandEffG1}
\end{multline}

The first term in the above is called the \textit{demand effect}, i.e., the direct effect of $a$ on $D_1$ and $\pi_1$. Note that this value can be either positive or negative based on the values of $a$ and $b$. Moreover, by using Equations~\ref{eq:Demand1}, and \ref{eq:PriceNEG}, we have:
\begin{multline}
\frac{\partial D_1}{\partial p_2} \frac{\partial p_2^*}{\partial a} = \left( \frac{1}{2T\left(1 - a - b\right) }\right) \left( T \left( \frac{-4+2a}{3} \right) \right) 
= \frac{a -2}{3\left(1 -a - b\right)},
\label{eq:StrategicEffG1}
\end{multline} 
which is called the \textit{strategic effect}. It shows the indirect effect through the change in firm 2's price on $D_1$ and $\pi_1$. Note that the above term is negative. By using Equations~\ref{eq:GLoc2}, \ref{eq:DemandEffG1} and \ref{eq:StrategicEffG1}, we have:
\begin{multline}
\frac{d \pi_1 }{d a} = p_1^* \left(\frac{-1 -3a -b}{6\left(1 - a -b \right)}  + \frac{\beta \left(q_1 - q_2\right)}{6T\left(1-a-b\right)^2}\right) \\
- 2C_1 \left(a - Z_A\right) - 2S_1 q_1^2 \left(a - Z_A\right).
\label{eq:GLoc3}
\end{multline}

Note that we restrict the values of $a$ to be in the interval $0 \leq a \leq Z_A $. By setting the above equation to zero, we find the corresponding critical points, and they also should satisfy condition $0 \leq a \leq Z_A$. In addition to these points, we should also check the boundary points, i.e., $0$ and $Z_A$. As a result, we have $\bar{a} = \{0, Z_A, a | 0\leq a \leq Z_A , \frac{d \pi_1}{da}=0\}$. This set introduces the potential solutions for vendor's 1 locations. Note that $\bar{a}$ and $opt_{q_1}\left(a,b,q_2\right)$ give the potential best responses. The one that gives the highest utility provides vendor 1's best response that is function of $b$ and $q_2$. 



In a similar way, for vendor 2 we have:
\begin{equation}
\frac{d \pi_2 }{d b} = \frac{\partial \pi_2 }{\partial p_2} \frac{\partial p_2}{\partial b} + \frac{\partial \pi_2 }{\partial b} + \frac{\partial \pi_2}{\partial p_1} \frac{\partial p_1}{\partial b}.
\label{eq:1Loc1}
\end{equation}

According to the envelope theorem and $\frac{\partial \pi_2}{\partial p_2} = 0$ in step 2, we have:
\begin{multline}
\frac{d \pi_2 }{d b} = \frac{\partial \pi_2 }{\partial b} + \frac{\partial \pi_2}{\partial p_1} \frac{\partial p^*_1}{\partial b} = 2C_2 \left( 1 - b -Z_A \right)  \\ + 2S_2q_2^2 \left( 1 - b -Z_A \right) + p_2^* \left( \frac{\partial D_2}{\partial b}  + \frac{\partial D_2}{\partial p_1} \frac{ \partial p_1^*}{\partial b}\right).
\label{eq:1Loc2}
\end{multline}

By using Equations~\ref{eq:DemandG}, \ref{eq:PriceNE1}, and \ref{eq:PriceNEG} the demand effect of vendor 2 is:

\begin{multline}
\frac{\partial D_2}{\partial b} = \frac{1}{2} + \frac{p_1^* - p_2^* + \beta\left(q_2 - q_1\right)}{2T\left(1 - a - b\right)^2}  
= \frac{3 - 5b - a}{6\left(1 - a - b\right)} + \frac{\beta \left(q_2-q_1\right)}{6T\left(1-a-b\right)^2}.
\label{eq:DemandEff1-1}
\end{multline}

The strategic effect of vendor 2 based on Equations~\ref{eq:DemandG} and \ref{eq:PriceNE1} is equal to:
\begin{multline}
\frac{\partial D_2}{\partial p_1} \frac{\partial p_1^*}{\partial b} = \left( \frac{1}{2T\left(1 - a - b\right) }\right) \left( T \left( \frac{-4+2b}{3} \right) \right) =
\frac{b -2}{3\left(1 -a - b\right)}.
\label{eq:StrategicEff1-1}
\end{multline} 

By using the demand effect and the strategic effect of vendor 2, we have:
\begin{multline}
\frac{d \pi_2}{d b} = p_2^* \left(\frac{-1 -3b -a}{6\left(1 - a -b \right)}  + \frac{\beta \left(q_2 - q_1\right)}{6T\left(1-a-b\right)^2}\right) \\
+ 2C_2 \left(1-b - Z_A\right) + 2S_2 q_2^2 \left(1-b - Z_A\right).
\label{eq:1Loc3}
\end{multline}

Note that we restrict the values of $b$ to be in interval $0 \leq b \leq Z_A $. By setting the above equation to zero, we find the corresponding critical points, and they also should satisfy condition $0 \leq b \leq Z_A$. In addition to these points, we should also check the boundary points, i.e., $0$ and $Z_A$. As a result, we have $\bar{b} = \{0, Z_A, b | 0\leq b \leq Z_A , \frac{d \pi_2}{db}=0\}$. This set introduces the potential solutions for vendor's 2 locations. Note that $\bar{b}$ and $opt_{q_2}\left(a,b,q_1\right)$ give the potential best responses. The one that gives the highest utility provides vendor 2's best response that is function of $a$ and $q_1$.

\subsection{Proof of Lemma~\ref{lem:QualInvest}}
\label{sub:SecInv}

According to a vendor's security quality best responses, e.g., Equation~\ref{eq:BRq1NoF}, best response security equality is not equal to zero unless $B_1$ is equal to zero and $A_1$ is negative. If $q_2=0$, $B_1$ is positive regardless of the location choices. Therefore, we have $opt_{q_1}\left(a,b,q_2\right) = Q$ or $opt_{q_1}\left(a,b,q_2\right) = \min\left(\frac{-B_1}{A_1},Q\right)$ which are not equal to zero. $q_1=q_2=0$ is NE if $B_1=B_2=0$ and $A_1,A_2 <0$ are both satisfied. $B_1=B_2=0$ never happens unless $1-a-b=0$ which means that both vendors are at the same location. If both vendors are at the same location as $Z_A$, this means that both of them have the quality $Q$. If they are at a different location than $Z_A$, the utility of customization is negative which is not acceptable for both vendors. Therefore, $q_1=q_2=0$ is not NE. $\qed$

\subsection{Price Nash Equilibrium with Fine}
\label{sub:proofPriceNE}
By taking the partial derivative with respect to price of firm 1, we have:

\begin{multline}
\frac{\partial \pi_1}{\partial p_1} = D_1 + \left(p_1 - f_1\right)  \frac{\partial D_1}{\partial p_1} = \frac{p_2 - p_1}{2T\left(1-a-b\right)} + a 
\\ 
+ \frac{1 -a -b}{2} + \frac{\beta \left(q_1-q_2\right)}{2T\left(1-a-b\right)} + \left(p_1 - f_1\right) \left(\frac{-1}{2T\left(1 -a - b\right)} \right).
\label{eq:PricePartial1F}
\end{multline}

The above equation is linear in $p_1$ and its coefficient is negative. Setting the above value to zero gives the following:
\begin{multline}
\frac{\partial \pi_1}{\partial p_1} = 0 \, \, \, \Rightarrow \, \, \, p^{crit}_1 =  \frac{p_2}{2} +   \frac{T}{2} \left(1 -a - b \right) \left(1 + a -b \right) 
+ \frac{\beta}{2} \left(q_1 - q_2\right) + \frac{f_1}{2}.
\label{eq:PricePartial1-2F}
\end{multline} 

Assuming that $p^{crit}_1 \geq 0$, $\pi_1$ is increasing in $[0, p_1^{crit}]$ and decreasing in $[p_1^{crit}, \infty]$. Therefore, $p_1^{crit}$ is maximum. 

In a similar way, for vendor 2 we have:

\begin{multline}
\frac{\partial \pi_2}{\partial p_2} =D_2 + \left(p_2 - f_2\right)  \frac{\partial D_2}{\partial p_2} = \frac{p_1 - p_2}{2T\left(1-a-b\right)} + b\\
 + \frac{1 -a -b}{2} + \frac{\beta \left(q_2 - q_1\right)}{2T\left(1-a-b\right)} + \left(p_2 - f_2\right) \left(\frac{-1}{2T\left(1 -a - b\right)} \right),
\label{eq:PricePartialGF}
\end{multline}

which is linear in $p_2$ and its coefficient is negative. By setting the above equation to zero, we have:
\begin{multline}
\frac{\partial \pi_2}{\partial p_2} = 0 \, \, \, \Rightarrow \, \, \, p^{crit}_2 =  \frac{p_1}{2} +   \frac{T}{2} \left(1 -a - b \right)\left(1-a+b\right) \\
+\frac{\beta}{2} \left(q_2 -q_1\right) + \frac{f_2}{2}.
\label{eq:PricePartialG-2F}
\end{multline}

Assuming that $p^{crit}_2 \geq 0$,$\pi_2$ is increasing in $[0, p_G^{crit}]$ and decreasing in $[p_G^{crit}, \infty]$. Hence, maximum is $p_2^{crit}$. 

Solving both Equations~\ref{eq:PricePartial1-2F} and \ref{eq:PricePartialG-2F} provides the Nash equilibrium in price being represented in Theorem~\ref{thm:PriceNEFine}. Note that this Nash equilibrium always exists. $\qed$

\subsection{Proof of Lemma~\ref{lem:Vend1BR} }
\label{sub:proofLocNo}

In order to calculate $a$ and $b$, based on Equations~\ref{eq:GLoc3} and \ref{eq:1Loc3}, we have:

\begin{equation}
\frac{d \pi_1 }{d a} = p_1^* \left(\frac{-1 -3a -b}{6\left(1 - a -b \right)}\right)- 2C_1 \left(a - Z_A\right),
\label{eq:GLoc3-NoSec}
\end{equation}

\begin{equation}
\frac{d \pi_2}{d b} = p_2^* \left(\frac{-1 -3b -a}{6\left(1 - a -b \right)}\right)
+ 2C_2 \left(1-b - Z_A\right),
\label{eq:1Loc3-NoSec}
\end{equation}

where $p_1^*$ and $p_2^*$ are equal to Equations~\ref{eq:PriceNE1-No} and \ref{eq:PriceNEG-No}, respectively. By assigning $p_1^*$ and $p_2^*$ into the above equations, we have:

\begin{multline}
\frac{d \pi_1 }{d a} = \frac{1}{18}
(-3Ta^2 + a \left(2Tb - 10T -36C_1\right) + T\left(b^2-2b-3\right)+36C_1Z_A),
\label{eq:GLoc3-NoSec2}
\end{multline}
which is quadratic in terms of $a$.

\begin{multline}
\frac{d \pi_2}{d b} = \frac{1}{18}
(-3Tb^2 + b \left(2Ta - 10T -36C_2\right) \\ + T\left(a^2-2a-3\right)+36C_2(1-Z_A),
\label{eq:1Loc3-NoSec2}
\end{multline}
which is quadratic in terms of $b$.

In a quadratic function like $Ax^2+Bx+C=0$, if $\Delta=B^2 - 4AC \geq 0$, that equation has two roots, which we call $x_1$ and $x_2$. Further, we have $x_1+x_2 = \frac{-B}{A}$ and $x_1x_2 = \frac{C}{A}$. If $\Delta < 0$, the quadratic function does not reach zero and its sign is the same as the sign of $A$. 

By setting Equation~\ref{eq:GLoc3-NoSec2}, i.e., $\frac{d\pi_1}{da} = 0$, we have a quadratic function in $a$. In this equation, we have $A=-3T < 0$, $B = 2Tb - 10T -36C_1 < 0 $, and $C = T\left(b^2-2b-3\right)+36C_1Z_A$. If $C \leq 0$, $\Delta$ can be both positive or negative. If $\Delta \leq 0$, this means that $\frac{d\pi_1}{da}$ is negative for any values of $a$. In other words, $\pi_1$ is decreasing in terms of $a$. Therefore, we have $a^*=0$. If $\Delta >0$, this means that $\frac{d\pi_1}{da} =0$ has two roots, i.e., $a_1$ and $a_2$. Moreover, we have $a_1a_2 = \frac{C}{A} > 0$ and $a_1 + a_2 = \frac{-B}{A} < 0$. This means that both $a_1$ and $a_2$ are negative and for any non-negative value of $a$, $\frac{d\pi_1}{da} $ is negative. This gives that $a^*=0$, In summary, when $C \leq 0$, we have $a^*=0$.

For the case where $C > 0$, we have $\Delta > 0$ and as result $\frac{d\pi_1}{da} = 0$ has two roots, i.e., $a_1$ and $a_2$. One of them is negative, $a_1 <0$, and the other one is positive, $a_2 > 0$. Because, we have $a_1a_2 = \frac{C}{A} < 0$ and $a_1 + a_2 = \frac{-B}{A} < 0$. Hence, $\pi_1$ is increasing in $[0,a_2]$. Note that we have assumed that $a \in [0, Z_A]$. Therefore, we have $a^* = \min\{a_2 , Z_A\}$. In summary, we have:

\[
a^* =
\begin{cases}
0 & \text{if $T\left(b^2-2b-3\right)+36C_1Z_A \leq 0$} \\
\min\{a_2,Z_A\} & \text{if $T\left(b^2-2b-3\right)+36C_1Z_A > 0$}. \\
\end{cases}
\]

The condition of the above equation is quadratic in $b$. Let's call the root of this quadratic function (if it exists), i.e., $T\left(b^2-2b-3\right)+36C_1Z_A = 0$, as $\bar{b}_1$ and $\bar{b}_2$, where $\bar{b}_1 \leq \bar{b}_2$. By doing a similar analysis, we can rewrite $a^*$ as follows:

\[
a^* =
\begin{cases}
0 & \text{if $C_1 \leq \frac{T}{12Z_A}$} \\
\min\{a_2,Z_A\} & \text{if $C_1 \geq \frac{T}{9Z_A}$} \\
\min\{a_2,Z_A\} & \text{if $\frac{T}{12Z_A} < C_1 < \frac{T}{9Z_A}$} \\
&  \text{$\&$ $0 \leq b \leq \min\{Z_A,\bar{b}_1\}$ }\\
0 & \text{if $\frac{T}{12Z_A} < C_1 < \frac{T}{9Z_A}$}\\
&  \text{$\&$ $\bar{b}_1 \leq Z_A$ $\&$ $\bar{b}_1 \leq b \leq Z_A$}, \\
\end{cases}
\]

where $\bar{b}_1 = 1 - \sqrt{4 - \frac{36C_1}{T}Z_A}$.

In a similar way, for vendor 2 we have:

\[
b^* =
\begin{cases}
0 & \text{if $T\left(a^2-2a-3\right)+36C_2(1-Z_A) \leq 0$} \\
\min\{b_2,Z_A\} & \text{if $T\left(a^2-2a-3\right)+36C_2(1-Z_A) > 0$},\\
\end{cases}
\]

where $b_1$ and $b_2$ are two roots of $\frac{d \pi_2}{db}= 0$ in which $b_1 \leq b_2$. Note that the condition of the above equation is quadratic in terms of $a$. Similar to vendor 1, we have:

\[
b^* =
\begin{cases}
0 & \text{if $C_2 \leq \frac{T}{12(1-Z_A)}$} \\
\min\{b_2,Z_A\} & \text{if $C_2 \geq \frac{T}{9(1-Z_A)}$} \\
\min\{b_2,Z_A\} & \text{if $\frac{T}{12(1-Z_A)} < C_2 < \frac{T}{9(1-Z_A)}$} \\
&  \text{$\&$ $0 \leq a \leq \min\{Z_A,\bar{a}_1\}$ }\\
0 & \text{if $\frac{T}{12(1-Z_A)} < C_2 < \frac{T}{9(1-Z_A)}$}\\
&  \text{$\&$ $\bar{a}_1 \leq Z_A$ $\&$ $\bar{a}_1 \leq a \leq Z_A$}, \\
\end{cases}
\]

where $\bar{a}_1 = 1 - \sqrt{4 - \frac{36C_1}{T}(1-Z_A)}$.

\subsection{Quality Nash Equilibrium with Fine}
\label{sub:qualityNEfine}

For quality calculation, we should take into account that the calculated price is a function of quality.
By taking the partial derivative of Equation~\ref{eq:venforUtilF} with respect to quality, we have:
\begin{equation}
\frac{d \pi_1}{d q_1} =  \frac{\partial \pi_1}{\partial q_1} + \frac{\partial \pi_1}{\partial p_1} \frac{\partial p_1}{\partial q_1} + \frac{\partial \pi_1}{\partial p_2} \frac{\partial p_2}{\partial q_1} .
\label{eq:QualPartial1F}
\end{equation}

Note that in the above equation, we are going to maximize $\pi_1$ given that Equation~\ref{eq:PriceNE1F} is satisfied. According to the envelope theorem, vendor 1 maximizes its utility with respect to the price in the second stage where $\frac{\partial \pi_1}{\partial p_1} = 0$. Therefore, we have:

\begin{multline}
\frac{d \pi_1}{d q_1} =  \frac{\partial \pi_1}{\partial q_1} + \frac{\partial \pi_1}{\partial p_2} \frac{\partial p_2}{\partial q_1} \, \, \, \Rightarrow \,\,\,\frac{d \pi_1}{d q_1}  =  -2S_1 q_1 \left(a - Z_A\right)^2  \\
-D_1 \frac{\partial f_1}{\partial q_1} + \left(p^*_1-f_1\right) \frac{\partial D_1}{\partial q_1} + p_1^* \frac{\partial D_1}{\partial p_2} \frac{\partial p_2^*}{\partial q_1}.
\label{eq:QualPartial2F}
\end{multline}

By using Equations~\ref{eq:Demand1} and \ref{eq:PriceNEGF}, we have:

\begin{equation}
\frac{\partial D_1}{\partial p_2} \frac{\partial p_2^*}{\partial q_1} = \left( \frac{1}{2T\left(1 - a - b\right)} \right) \left(\frac{1}{3} \frac{\partial f_1}{\partial q_1} - \frac{\beta}{3}\right).
\label{eq:QualPart1F}
\end{equation}
Both the above equations show that we need to calculate $\frac{\partial f_1}{\partial q_1}$. By considering Equation~\ref{eq:Fine}, we have: 

\begin{equation}
\frac{\partial f_1}{\partial q_1} = \begin{cases}
- F & \text{if } q^{min} \geq q_1\\
0              & \text{if } q^{min} < q_1.
\end{cases}
\label{eq:FineDer}
\end{equation}

Note that for $q_1 = q^{min}$, $\frac{\partial f_1}{\partial q_1}$ is not defined. But, here, we assume that this value is equal to $-F$. Therefore, for $q_1 \leq q^{min}$, Equation~\ref{eq:QualPartial2F} changes into:

\begin{multline}
\frac{\partial \pi_1}{\partial q_1} =   q_1 \left( -2S_1\left(a - Z_A\right)^2 + \frac{\left(\beta+ F\right)^2}{9T\left(1-a-b\right)} \right) \\
+ \frac{\beta + F}{9} \left( 3+a-b + \frac{f_2 -\beta q_2 - F q^{min}}{T\left( 1- a - b \right)} \right).
\label{eq:QualPartial3F}
\end{multline}

The above equation is linear in $q_1$. In a linear equation like $cq_1 + d$ in which $c,d \geq 0$, $cq_1 + d$ is positive for every nonnegative value of $q_1$. If this condition is satisfied for Equation~\ref{eq:QualPartial3F}, $\pi_1$ is increasing in $[0,q^{min}]$ and as a result, $\pi_1\left(a,b,q\right)$ is maximized at $q^{min}$ in interval $[0,q^{min}]$. Here, in our analysis, we assume that the consumers do not take security into account, i.e., $\beta=0$.  
Hence, for $q_1 > q^{min}$, by considering Equations~\ref{eq:Fine} and \ref{eq:FineDer}, Equation~\ref{eq:QualPartial2F} is equal to the following:

\begin{equation}
\frac{\partial \pi_1}{\partial q_1} = - 2S_1q_1\left( a - Z_A \right) ^2 .
\label{eq:QualPartial4F}
\end{equation}

The above Equation is negative for all values of $q_1$. This means that $\pi_1$ is decreasing in $q_1$ and as a result, $\pi_1$ is maximized at $q^{min}$ in interval $[q^{min}, Q]$. Therefore, by considering both Equations~\ref{eq:QualPartial3F} and \ref{eq:QualPartial4F}, we have $q^*_1 = q^{min}$ if the following two conditions are satisfied:

\begin{equation}
F^2 - 18TS_1 \left(1 - a - b \right)(a - Z_A)^2 \geq 0,
\label{eq:QualCond1-App}
\end{equation}

\begin{equation}
3+a-b + \frac{f_2 - F q^{min}}{T\left( 1- a - b \right)} \geq 0 .
\label{eq:QualCond2-App}
\end{equation}

In a similar way, vendor 2 invests in $q_2^* = q^{min}$ if the following two conditions are satisfied: 

\begin{equation}
F^2 - 18TS_2 \left(1 - a - b \right)(1- b - Z_A)^2 \geq 0,
\label{eq:QualCond1-AppV2}
\end{equation}

\begin{equation}
3+a-b + \frac{f_1 - F q^{min}}{T\left( 1- a - b \right)} \geq 0 .
\label{eq:QualCond2-AppV2}
\end{equation}

If both vendor 1 and vendor 2 invest in $q^{min}$ level of security, we have $f_1=f_2=0$. If Equations~\ref{eq:QualCond1-App} and \ref{eq:QualCond1-AppV2} are satisfied in addition to the following equation, both vendors will invest in $q^{min}$ level of security.

\begin{equation}
3+a-b - \frac{ F q^{min}}{T\left( 1- a - b \right)} \geq 0 . \qed
\label{eq:QualCond2-AppV}
\end{equation}

\subsection{Proof of Lemmas~\ref{lem:Vend1BRFine}}
\label{proof:qualFine}

When consumers do not take security into account, i.e., $\beta=0$, and $q_1^*=q_2^*=q^{min}$, we have: 

\begin{equation}
\frac{d \pi_1 }{d a} = p_1^* \left(\frac{-1 -3a -b}{6\left(1 - a -b \right)}\right)- 2\left(C_1 +S_1 \left(q^{min}\right)^2\right) \left(a - Z_A\right),
\label{eq:GLoc3-NoSecF}
\end{equation}

\begin{multline}
\frac{d \pi_2}{d b} = p_2^* \left(\frac{-1 -3b -a}{6\left(1 - a -b \right)}\right)
+ 2\left(C_2 +S_2 \left(q^{min}\right)^2\right) \left(1-b - Z_A\right).
\label{eq:1Loc3-NoSecF}
\end{multline}

It is similar to the case where there is no fine except that $C_1$ and $C_2$ change into $C_1 +S_1 \left(q^{min}\right)^2$ and $C_2 +S_2 \left(q^{min}\right)^2$, respectively. $\qed$

\end{document}